\documentstyle[epsfig]{mn2e}
\begin{document}

\title[Cosmology with photometric redshift surveys]{Cosmology with
  photometric redshift surveys}

\author[Chris Blake \& Sarah Bridle]{Chris Blake$^{1,2,}$\footnotemark
  \, \& Sarah Bridle$^{3,}$\footnotemark \\ \\ $^1$ Department of
  Physics \& Astronomy, University of British Columbia, 6224
  Agricultural Road, Vancouver, B.C., V6T 1Z1, Canada \\ $^2$ School
  of Physics, University of New South Wales, Sydney, NSW 2052,
  Australia \\ $^3$ Department of Physics \& Astronomy, University
  College London, London, WC1E 6BT, U.K.}

\maketitle

\begin{abstract}
We explore the utility of future photometric redshift imaging surveys
for delineating the large-scale structure of the Universe, and assess
the resulting constraints on the cosmological model.  We perform two
complementary types of analysis:

(1) We quantify the statistical confidence and the accuracy with which
such surveys will be able to detect and measure characteristic
features in the clustering power spectrum such as the {\it acoustic
  oscillations} and the {\it turnover}, in a `model-independent'
fashion.  We show for example that a $10{,}000$ deg$^2$ imaging survey
with depth $r = 22.5$ and photometric redshift accuracy $\delta
z/(1+z) = 0.03$ will detect the acoustic oscillations with $99.9$ per
cent confidence, measuring the associated preferred cosmological scale
with $2$ per cent precision.  Such a survey will also detect the
turnover with $95$ per cent confidence, determining the corresponding
scale with $20$ per cent accuracy.

(2) By assuming a $\Lambda$CDM model power spectrum we calculate the
confidence with which a non-zero baryon fraction can be deduced from
such future galaxy surveys.  We quantify `wiggle detection' by
calculating the number of standard deviations by which the baryon
fraction is measured, after marginalizing over the shape parameter.
This is typically a factor of four more significant (in terms of
number of standard deviations) than the above `model-independent'
result.

For both analyses we quantify the variation of the results with
magnitude depth and photometric redshift precision, and discuss the
prospects for obtaining the required performance with realistic future
surveys.  We conclude that the precision with which the clustering
pattern may be inferred from future photometric redshift surveys will
be competitive with contemporaneous spectroscopic redshift surveys,
assuming that systematic effects can be controlled.  We find that for
equivalent wiggle-detection power, a photometric redshift survey
requires an area approximately $12(\delta z/(1+z)/0.03)$ times larger
than a spectroscopic survey, for a given magnitude limit.  We also
note that an analysis of Luminous Red Galaxies in the Sloan Digital
Sky Survey may yield a marginal detection of acoustic oscillations in
the imaging survey, in addition to that recently reported for the
spectroscopic component.

\end{abstract}

\section{Introduction}
\renewcommand{\thefootnote}{\fnsymbol{footnote}}
\setcounter{footnote}{1}
\footnotetext{E-mail: cab@astro.ubc.ca}
\setcounter{footnote}{2}
\footnotetext{E-mail: sarah@sarahbridle.net}

Today's most pressing cosmological questions demand the construction
of galaxy surveys of unprecedented depth and volume.  Such questions
include: is the accelerating rate of cosmic expansion driven by
Einstein's cosmological constant or some different form of `dark
energy'?  What are the properties of this dark energy?  Can competing
models of inflation be discriminated by accurate measurements of the
shape of the primordial power spectrum of mass fluctuations?

Galaxy surveys delineate the large-scale structure of the Universe and
thereby provide a powerful and independent constraint on the
cosmological model.  The currently-favoured `concordance model' -- in
which $\approx 70$ per cent of the energy density of today's Universe
is resident in a relatively unclustered form known as `dark energy' --
is evidenced by a combination of observations of the Cosmic Microwave
Background (e.g.\ Spergel et al.\ 2003) with either those of galaxy
clustering (e.g.\ Percival et al.\ 2001) or of high-redshift
supernovae (e.g.\ Riess et al.\ 1998; Perlmutter et al.\ 1999).
Either pair of these independent datasets are required to break the
degeneracies between model parameters and render a unique cosmology.

According to standard cosmological theory, if the linear-regime
clustering power spectrum is measured with sufficient precision then
it will no longer appear smooth and monotonic: specific features and
modulations will become apparent.  Two such attributes are predicted:
firstly, a series of {\it acoustic oscillations} -- sinusoidal
modulations in power as a function of scale imprinted in the baryonic
component before recombination (Peebles \& Yu 1970; Hu \& Sugiyama
1996) -- and secondly, a {\it turnover} -- a broad maximum in
clustering power on large scales originating from the
radiation-dominated epoch.  These features encode characteristic
cosmological scales that can be extracted from the observations,
greatly improving constraints upon cosmological models (e.g.\ Blake \&
Glazebrook 2003; Seo \& Eisenstein 2003).  Moreover, other
currently-unknown modulations in power (e.g.\ signatures of inflation)
may be discovered when the clustering pattern is examined with
sufficiently high precision (e.g.\ Martin \& Ringeval 2004).

Very recently, the acoustic signature has been convincingly identified
for the first time in the clustering pattern of Luminous Red Galaxies
in the Sloan Digital Sky Survey (SDSS; Eisenstein et al.\ 2005).  The
2dF Galaxy Redshift Survey has produced consistent measurements (Cole
et al.\ 2005).  These results confirm previous tantalizing hints
(e.g.\ Miller, Nichol \& Chen 2002; Percival et al.\ 2001).  The
challenge now is to make more accurate measurements at different
redshifts, using these features to further constrain the cosmological
parameters, in particular the dark energy model.  At low redshift the
available volume is limited: the effect of {\it cosmic variance} is
significant.  Therefore such surveys are insensitive to clustering
modes on very large scales and are hampered by non-linear growth of
structure on small scales.  Higher-redshift large-scale surveys are
consequently required to map greater cosmic volumes: tracing
clustering modes with longer wavelengths and additionally unveiling
the pattern of linear clustering to significantly smaller scales.

The high-redshift spectroscopic surveys currently being executed
(e.g.\ DEEP2, Davis et al.\ 2003; VVDS, Le Fevre et al.\ 2003) cover
solid angles of $\sim 10$ deg$^2$, which are insufficient for
detecting the predicted features in the clustering power
spectrum. Such projects are fundamentally limited by existing
instrumentation, being performed by spectrographs with relatively
small fields-of-view ($\approx 10 - 20'$) and restricted (albeit
impressive) multi-object capabilities. Some proposed new
instrumentation addresses this difficulty (e.g.\ the KAOS project,
Barden et al.\ 2004), permitting spectroscopic exposures over single
fields of $\approx 1$ deg$^2$ using $\approx 5000$ fibres.  However,
these projects will take many years to reach completion.

In this paper we consider the role that {\it photometric redshift
  catalogues derived from deep imaging surveys} could play in
addressing the scientific goals outlined above.  Extensive imaging
surveys (covering $\sim 10{,}000$ deg$^2$) to reasonable depths ($r
\approx 22$) are ongoing (e.g.\ SDSS); the implied redshift
distributions map galaxies over cosmic distances to $z \approx 1$ with
sufficient number density that clustering measurements are limited by
cosmic variance rather than by shot noise.  Future deeper imaging
surveys (e.g.\ Pan-STARRS, Kaiser et al.\ 2000; CTIO Dark Energy
Survey, {\tt http://cosmology.astro.uiuc.edu/DES}; LSST, Tyson et
al.\ 2002) are being planned to address a host of scientific questions
including in particular weak gravitational lensing.  We argue that
such surveys will also provide powerful measurements of features in
the galaxy clustering pattern.

The utility of photometric redshifts -- derived from broadband galaxy
colours rather than from spectra -- has been well-established, with
many different techniques being successfully utilized.  The simplest
method involves the fitting of model spectral templates
(e.g.\ Bolzonella, Miralles \& Pello 2000).  Other approaches use
spectroscopic `training sets' to calibrate the photometric redshifts
via an empirical polynomial of colour terms (Connolly et al.\ 1995) or
an artificial neural network (Firth, Lahav \& Somerville 2003).  The
precision $\delta z$ with which galaxy redshifts (and therefore radial
distances) may be determined varies with the method and filter set
used, together with the galaxy type, magnitude and redshift, but at
best is currently $\sigma_0 \equiv \delta z / (1+z) \sim 0.03$
(e.g.\ COMBO-17, Wolf et al.\ 2003).  For the SDSS imaging component,
the rms photometric redshift accuracy varies from $\delta z \approx
0.03$ for bright galaxies with $r < 18$ to $\delta z \approx 0.1$ for
magnitudes $r \approx 21$ (Csabai et al.\ 2003).

The blurring of large-scale structure in the radial direction due to
the photometric redshift error degrades measurements of the clustering
pattern.  However, {\it on physical scales larger than that implied by
  the redshift error, the information is preserved}.  Moreover, on
smaller scales the tangential information always survives, and the
vast area which may be readily covered by an imaging survey can
potentially provide more independent structure modes on a given scale
than those yielded by a fully spectroscopic survey of a smaller solid
angle, implying very competitive cosmological constraints.
Photometric redshifts have already been used to construct
volume-limited samples of low-redshift galaxies and measure their
angular clustering properties (Budavari et al.\ 2003; see also
Meiksin, White \& Peacock 1999, Cooray et al.\ 2001).  The
cosmological parameter constraints resulting from future photometric
redshift imaging surveys have been simulated by Seo \& Eisenstein
(2003); Amendola, Quercellini \& Giallongo (2004) and Dolney, Jain \&
Takada (2004).

In this study we use a Monte Carlo approach to model the galaxy power
spectra resulting from a host of simulated photometric redshift
surveys as a function of the limiting magnitude of the initial imaging
and the accuracy of the derived photometric redshift.  Our simulation
methodology is described in Section \ref{secmeth}, where first results
for the accuracy of power spectrum measurements are presented.  We
infer constraints on the cosmological model using two complementary
methods with very different prior assumptions. Firstly, in Sections
\ref{secwig} and \ref{secturn} we discuss in detail the resulting
confidence of detection of the acoustic oscillations and power
spectrum turnover, respectively, and the accuracy with which the
associated characteristic cosmological scales may be extracted.  In
these analyses we make minimal assumptions, purely concerning
ourselves with the statistical detection of power spectrum features
relative to a smooth monotonic fit.  Secondly, in Section
\ref{seccospar} we use the full power spectrum shape information in
conjunction with theoretical fitting formulae to compute constraints
on the basic parameters of the cosmological model, in particular the
baryon fraction $\Omega_{\rm b}/\Omega_{\rm m}$ and the running of the
spectral index of the primordial power spectrum $n_{\rm run}$. In all
cases we compare our results to those deduced from spectroscopic
redshift surveys. We evaluate the effect of the approximations of our
methodology in Section \ref{secapprox}, in particular considering a
wider range of photometric redshift error distributions. Finally, in
Section \ref{secrealsurv} we outline the prospects for obtaining the
requisite imaging depth and photometric redshift accuracy using
realistic future surveys.

\section{Modelling the power spectrum of photometric redshift surveys}
\label{secmeth}

\subsection{Method summary}

Our methodology for simulating the large-scale structure of future
galaxy surveys is to generate many `Monte Carlo' realizations of
galaxy distributions from a fiducial power spectrum.  An observed
power spectrum is measured for each realization separately, using
techniques similar to those which would be employed for real survey
data.  The resulting ensemble of observed power spectra can then be
used to quantify the error distribution in derived quantities, without
any need to approximate the likelihood surface by techniques such as
Fisher matrices (e.g.\ Seo \& Eisenstein 2003; Amendola et al.\ 2004;
Dolney et al.\ 2004).  For example, the standard deviation in the
measurement of the power spectrum $P(k)$ in a given bin around scale
$k$ follows from the scatter in the recovered values of $P(k)$ in that
bin across the realizations, without the need for analytic
approximations.

Our procedure for modelling photometric redshifts is to convolve the
simulated galaxy distributions in the radial direction with a
photometric redshift error function (in the fiducial case, a Gaussian
with width $\sigma_x$ in real space). The measured power spectrum is
derived by computing the Fourier transform of the whole survey box,
then discarding small-scale radial Fourier modes with wavenumbers
$k_{\rm rad} \ga 1/\sigma_x$ (which contribute no signal due to the
radial smearing). Note that the resulting number of useful Fourier
structure modes is identical to that obtained if the survey box is
instead split into many independent slabs of width $\sigma_x$ and a
purely {\it angular} power spectrum is measured for each slice.

In particular, we wish to assess the confidence with which we can
detect specific features in the clustering power spectrum such as the
acoustic oscillations and the `turnover'.  These features can be
modelled by simple empirical formulae which can be fitted to each
measured power spectrum realization, and the best-fit $\chi^2$
statistic calculated.  The resulting best-fit $\chi^2$ can be compared
with that of a smooth (featureless) power spectrum fit, resulting in a
{\it relative probability of feature detection for each realization}.
The distribution of the relative probabilities across the Monte Carlo
realizations enables a very realistic assessment of the efficacy of
future surveys across the statistical ensemble of possible universes.

Furthermore, we are interested in recovering {\it characteristic
scales} from these features in the power spectrum.  These scales can
be encoded into our empirical fitting formulae; the distribution of
best-fitting values of these scales across the realizations is
indicative of the realistic accuracy with which it is possible to
measure them with the simulated survey.

In order to perform our simulations we must also adopt a fiducial set
of cosmological parameters, which determine both the cosmic volume
mapped by a given survey and the fiducial power spectrum (via the
fitting formulae of Eisenstein \& Hu 1998).

We characterize a photometric redshift imaging survey using two
parameters:
\begin{itemize}
\item The photometric redshift error distribution (in the simplest
  case, a Gaussian distribution with standard deviation $\delta z$),
  which controls the `smearing' of the underlying large-scale
  structure in the radial direction.
\item The threshold apparent magnitude of the input imaging catalogue,
  $m_{\rm lim}$, which determines the redshift distribution $dN/dz$ of
  the `unsmeared' galaxy distribution, i.e.\ the radial depth of the
  survey.  This magnitude limit is defined using the SDSS $r$-band
  filter.
\end{itemize}
In this paper we present the results of simulations of a range of
photometric redshift surveys as a function of these two parameters.
We assume the survey area in all cases is $10{,}000$ deg$^2$.

In the following Section we provide a detailed account of the
assumptions and method we used to simulate the observed power spectra.
For analyses of future spectroscopic redshift surveys using a similar
method we refer the reader to Blake \& Glazebrook (2003) and
Glazebrook \& Blake (2005).

\subsection{Detailed simulation methodology}
\label{secpksteps}

\begin{enumerate}

\item A fiducial cosmology is chosen for the simulation.  Unless
  otherwise stated we assumed a flat $\Lambda$CDM Universe with
  parameters $\Omega_{\rm m} = 0.3$, $\Omega_\Lambda = 0.7$, $h = 0.7$
  and $\Omega_{\rm b}/\Omega_{\rm m} = 0.15$.

\item A limiting apparent magnitude $m_{\rm lim}$ is assumed for the
  imaging survey.  A model redshift distribution $dN/dz(m_{\rm lim})$
  is determined, as described in Section \ref{secnz}.

\item A survey redshift range $(z_{\rm min},z_{\rm max})$ and solid
  angle $A_\Omega$ is specified.  We assumed $A_\Omega = 10{,}000$
  deg$^2$ for our imaging surveys.  The chosen redshift interval
  depends on $m_{\rm lim}$ as described in Section \ref{secnz} (and
  listed in Table \ref{tabsurv}).

\item We performed our simulations using a `flat-sky approximation'
  for computational convenience (this approximation has a negligible
  effect on our results as discussed in Section \ref{secapprox}).  A
  cuboid with sides of co-moving lengths $(L_x,L_y,L_z)$ is created,
  possessing a volume equal to that enclosed by the survey cone.  We
  take the $x$-axis as the radial direction. The length $L_x$ is the
  co-moving distance between redshifts $z_{\rm min}$ and $z_{\rm
    max}$, and the other dimensions are determined by stipulating $L_y
  = L_z$ (although the results are independent of the ratio $L_y/L_z$,
  assuming that both of these dimensions are large enough to imply a
  sensitivity to structural modes with scales contributing to the
  acoustic oscillations).

\item A model linear theory matter power spectrum $P_{\rm
  mass}(k,z=0)$ is computed for the chosen parameters $(\Omega_{\rm
  m},\Omega_{\rm b},h)$ from the fitting formula of Eisenstein \& Hu
  (1998), assuming a $z=0$ normalization $\sigma_8=1$ and a primordial
  power-law slope $n=1$.  The survey is assumed to have an `effective'
  redshift $z_{\rm eff} = (z_{\rm min} + z_{\rm max})/2$.  The power
  spectrum is scaled to this redshift using a standard $\Lambda$CDM
  growth factor:
\begin{equation}
P_{\rm gal}(k,z_{\rm eff}) = P_{\rm mass}(k,0) \, D_1(z_{\rm eff})^2 \, b^2
\label{eqpk}
\end{equation}
where we use the Carroll, Press \& Turner (1992) approximation for
$D_1(z)$ and a constant linear bias factor $b$ for the clustering of
galaxies with respect to matter.  The value $b = 1$ is assumed for our
surveys, unless otherwise stated.  We neglect any evolution of
clustering across the depth of the survey box.  This approximation is
discussed in Section \ref{secapprox}; we note that the clustering
amplitude of galaxies is known to evolve much less rapidly with
redshift than that of the underlying mass fluctuations (indicating an
evolution of the galaxy bias parameter in the opposite sense).

\item The location in $k$-space of the transition between the linear
  and non-linear regimes of gravitational clustering, $k_{\rm lin}$,
  is determined from $P_{\rm gal}(k)$ in a conservative manner (see
  Blake \& Glazebrook 2003, Figure 1).  We only measure power spectra
  in the linear regime, i.e.\ for scales $k < k_{\rm lin}$.

\item A set of Monte Carlo realizations (numbering 400 for all
  simulations presented here) is then performed to generate many
  different galaxy distributions consistent with $P_{\rm gal}(k)$, as
  described in steps (viii) and (ix).

\item A cuboid of Fourier coefficients is constructed with grid lines
  set by $dk_i=2\pi/L_i$, with a Gaussian distribution of amplitudes
  determined from $P_{\rm gal}(k)$, and with randomized phases.  The
  gridding is sufficiently fine that the Nyquist frequencies in all
  directions are significantly greater than $k_{\rm lin}$.

\item The Fourier cuboid is FFTed to determine the density field in
  the real-space box.  The result is modulated by the survey redshift
  distribution $dN/dz$.  This observed density field is then Poisson
  sampled to determine the number of galaxies in each grid cell.

\item A photometric redshift error distribution is assumed.  For our
  main set of simulations, we modelled this function as a Gaussian
  distribution, such that the radial co-moving co-ordinate $x$ of each
  galaxy was smeared by an amount $\delta x$ sampled from a
  probability distribution
\begin{equation}
f(\delta x) \propto \exp{ \left[ - \frac{1}{2} \left( \frac{\delta
x}{\sigma_x} \right)^2 \right] }.
\label{eqzerr}
\end{equation}
In practice we specified a redshift error parameter $\sigma_0$ and
derived $\sigma_x$ in accordance with the equation
\begin{equation}
\sigma_x = \delta z \frac{dx}{dz}(z = z_{\rm eff}) = \sigma_0 (1 + z_{\rm eff})
\frac{c}{H(z_{\rm eff})}
\label{eqsigx}
\end{equation}
where $H(z)$ is the value of the Hubble constant measured by an
observer at redshift $z$.  Equation \ref{eqsigx} encodes the expected
zeroth-order dependence of photometric redshift precision, $\delta z
\propto (1+z)$, originating from the stretching of galaxy spectra with
redshift for a filter system with constant spectral resolution
$\Delta\lambda/\lambda$.  We assess the effect of more complex
photometric redshift error distributions than Equation \ref{eqzerr} in
Section \ref{secphotozerr}.

\item The galaxy number distribution is `smeared' along the $x-$
  (radial) direction in accordance with the photometric redshift error
  function, taking pixelization effects into account.  The resulting
  distribution is our simulated photometric redshift survey.  We note
  that our simple photometric-redshift error model represents a
  convolution of the `unsmeared' galaxy number distribution with the
  error function $f(x)$ (Equation \ref{eqzerr}).  In this case,
  according to the convolution theorem, the resulting power spectrum
  signal is damped along the radial direction:
\begin{equation}
P(k_x,k_y,k_z) \rightarrow P(k_x,k_y,k_z) \times \exp{ [ - (k_x
\sigma_x)^2 ] }
\label{eqpkdamp}
\end{equation}
where $g(k_x) = \exp{[-(k_x \sigma_x)^2]}$ is the square of the
Fourier transform of $f(x)$.

\item The power spectrum of the resulting distribution is measured
  using standard estimation tools: essentially this involves taking
  the Fourier transform of the density field, subtracting that of the
  survey window function, and binning up the resulting modes in
  $k$-space (see e.g.\ Hoyle et al.\ 2002; note that we do not use the
  optimal-weighting technique presented by Feldman, Kaiser \& Peacock
  (1994) because this does not represent a simple convolution of the
  density field and consequently step (xiii) below would not be
  possible).  Power spectrum modes in Fourier space are divided into
  bins of total $k$, up to a maximum of $k_{\rm lin}$.  We only
  include modes with a value of $|k_x|$ less than a maximum $k_{x,{\rm
      max}}$.  This is because in accordance with Equation
  \ref{eqpkdamp}, the photometric-redshift smearing damps the
  clustering signal along the radial direction such that modes with
  high values of $|k_x|$ contribute only noise.  Hence the dominant
  contribution to power spectrum bins with $k > k_{x,{\rm max}}$
  originates from tangential Fourier modes with $k_x \approx 0$ and
  $\sqrt{k_y^2 + k_z^2} \approx k$.  The value of $k_{x,{\rm max}}$ is
  determined by the equation
\begin{equation}
k_{x,{\rm max}} = 2/\sigma_x
\label{eqkxmax}
\end{equation}
where the coefficient of 2 was determined by experiment to be optimal
for the surveys presented here.  Use instead of a coefficient of $1.5$
does not change the results significantly, but $1.0$ is sub-optimal.

\item The measured power spectrum $P(k)$ is `undamped' by dividing by
  a function $f_{\rm damp}(k)$.  This `damping function' was
  determined by binning the expression $\exp{[-(k_x \sigma_x)^2]}$
  (from Equation \ref{eqpkdamp}) as a function of total $k$ as
  described in step (xii).

\item An error bar is assigned to each power spectrum bin using the
  variance measured over the Monte Carlo realizations.

\end{enumerate}

We do not incorporate {\it redshift-space distortions} into our
simulations because the implied radial smearing due to peculiar
velocities is much less than that due to the photometric redshift
error.

\begin{figure*}
\center
\epsfig{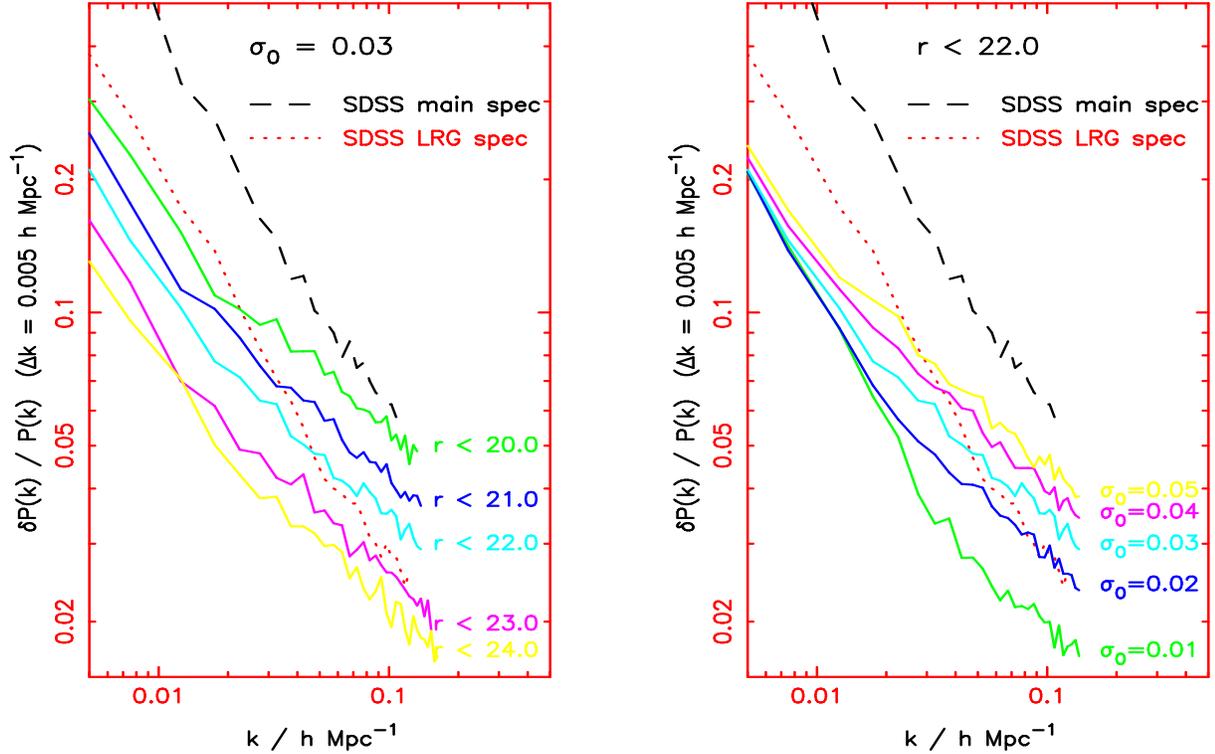}
\caption{Fractional error in $P(k)$ against $k$ for examples of
  photometric redshift imaging surveys.  The left-hand panel
  illustrates the variation of the fractional power spectrum error as
  a function of $r_{\rm lim}$ for $\sigma_0 = 0.03$; the right-hand
  panel displays the dependence on $\sigma_0$ for $r_{\rm lim} = 22$.
  We bin the power spectra in intervals of $\Delta k = 0.005 \, h$
  Mpc$^{-1}$.  Simulated results for the SDSS main spectroscopic
  survey (dashed line) and SDSS LRG spectroscopic survey (dotted line)
  are overplotted (see Section \ref{secpkerr} for details of our
  models of these surveys).  Considering the photometric redshift
  surveys, the fractional power spectrum precision improves with
  increasing survey depth (left-hand panel) owing to the greater
  cosmic volume mapped and the consequently higher density-of-states
  in $k$-space.  The performance degrades with increasing $\sigma_0$
  (right-hand panel) owing to the decreasing width of the `undamped'
  slab in $k$-space (i.e.\ the decreasing value of $k_{x,{\rm max}}$),
  although on very large scales ($k < k_{x,{\rm max}}$) all Fourier
  modes are retained and performance is unaffected.}
\label{figpksurv}
\end{figure*}

\subsection{Modelling the redshift distribution}
\label{secnz}

In order to model the `unsmeared' survey redshift distribution as a
function of the limiting apparent magnitude of the imaging survey
$m_{\rm lim}$, we used the luminosity functions derived from the
COMBO-17 survey (Wolf et al.\ 2003).  Table A.2 of Wolf et al.\ (2003)
lists Schechter function parameters in redshift slices of width
$\Delta z = 0.2$ in the range $0.2 < z < 1.2$ for the SDSS $r$
filter. In the regime $z < 0.2$ we applied the locally-determined SDSS
luminosity function (Blanton et al.\ 2003).  For a given threshold
apparent magnitude $r_{\rm lim}$, we converted these luminosity
functions into a redshift distribution $dN/dz$ for each redshift
slice, fitting the overall result with a simple model parameterized by
a characteristic redshift $z_0$ and an overall surface density
$\Sigma_0$ (in deg$^{-2}$):
\begin{equation}
\frac{dN}{dz} = \Sigma_0 \times \frac{3 z^2}{2 z_0^3} \exp{ \left[ -
\left( \frac{z}{z_0} \right)^{3/2} \right] }
\label{eqdndz}
\end{equation}
(e.g.\ Baugh \& Efstathiou 1993).  The values of the fitted parameters
are displayed in Table \ref{tabsurv}.  We used model K-corrections
averaged over different galaxy spectral types.

For each apparent magnitude limit we selected a redshift interval
$(z_{\rm min},z_{\rm max})$ for the simulation.  For all but the
shallowest surveys we set $z_{\rm min} = 0.2$, the results are
insensitive to this choice because there is minimal volume contained
by lower redshifts.  As the value of $z_{\rm max}$ increases, the
variance in the recovered power spectrum is determined by a balance
between two conflicting effects: increasing survey volume
(i.e.\ decreasing cosmic variance) and decreasing average number
density (i.e.\ increasing shot noise) owing to the fixed magnitude
threshold.  We determined the optimal value of $z_{\rm max}$ for each
magnitude threshold by experimenting to determine the most accurate
measurement of the acoustic oscillations (see Section
\ref{secwig}). Our chosen ranges are listed in Table \ref{tabsurv}.

We note that the optimal value of $z_{\rm max}$ for measuring power
spectrum modes around the turnover is marginally higher than that for
detecting the acoustic oscillations, because in the former case the
power spectrum amplitude is at a maximum, implying a lower required
galaxy number density for suppressing shot noise.  We always use the
more conservative values of $z_{\rm max}$ in Table \ref{tabsurv}, but
this does not change our results significantly.

\begin{table}
\center
\caption{Input parameters for simulated galaxy redshift surveys as a
  function of limiting apparent magnitude in the SDSS $r$ filter,
  $r_{\rm lim}$.  The `unsmeared' redshift distribution is specified
  by the values of $z_0$ and $\Sigma_0$ in accordance with Equation
  \ref{eqdndz}.  The minimum and maximum redshifts of the simulated
  survey, $z_{\rm min}$ and $z_{\rm max}$, are also listed.}
\label{tabsurv}
\begin{tabular}{ccccc}
\hline $r_{\rm lim}$ & $z_0$ & $\Sigma_0$ (deg$^{-2}$) & $z_{\rm min}$
& $z_{\rm max}$ \\
\hline
18.0 & 0.1 & 120 & 0.1 & 0.4 \\
18.5 & 0.12 & 230 & 0.1 & 0.5 \\
19.0 & 0.14 & 410 & 0.1 & 0.6 \\
19.5 & 0.16 & 710 & 0.1 & 0.6 \\
20.0 & 0.18 & 1200 & 0.2 & 0.7 \\
20.5 & 0.2 & 2000 & 0.2 & 0.7 \\
21.0 & 0.22 & 3200 & 0.2 & 0.8 \\
21.5 & 0.24 & 4900 & 0.2 & 0.8 \\
22.0 & 0.27 & 7500 & 0.2 & 0.9 \\
22.5 & 0.3 & 11100 & 0.2 & 1.0 \\
23.0 & 0.33 & 16300 & 0.2 & 1.2 \\
23.5 & 0.36 & 24000 & 0.2 & 1.3 \\
24.0 & 0.39 & 35400 & 0.2 & 1.4 \\
\hline
\end{tabular}
\end{table}

We note that all magnitudes referred to here are total galaxy
magnitudes, and we neglect all incompleteness effects associated with
surface brightness.  These are expected to be minimal at high redshift
owing to the decreasing apparent size of galaxies.  We also neglect
the fact that imaging surveys to depths $r_{\rm lim} \ga 23$ may be
used to select additional galaxy populations at $z \approx 3$ using
the Lyman Break technique.

\subsection{Fractional errors in the power spectrum}
\label{secpkerr}

We simulated photometric redshift surveys with limiting magnitudes
varying from $r_{\rm lim} = 20$ to $24$ and photometric redshift error
parameters in the interval $\sigma_0 = 0.01$ to $0.05$.  These ranges
were chosen to encompass the current state-of-the-art and realistic
future improvements.  Achieving precision $\sigma_0 = 0.01$ would
require observations with many narrow-band filters, probably
encompassing the near infra-red.

As a first application of our method, Figure \ref{figpksurv} displays
the fractional standard deviation in the power spectrum measurement as
a function of scale (averaged over the Monte Carlo realizations) for
some example photometric redshift surveys.  The measured power
spectrum modes are averaged over angles in bins of width $\Delta k =
0.005 \, h$ Mpc$^{-1}$.

The scaling of the resulting power spectrum errors as a function of
$k$ can be understood simply by counting the number of Fourier modes
$m$ within each bin (the errors $\delta P$ scaling as $1/\sqrt{m}$).
For a bin with $k \gg k_{x,{\rm max}}$, these modes are located
approximately within an cylindrical annulus in Fourier space of radius
$k$, thickness $\Delta k$ and depth $k_{x,{\rm max}}$.  This amounts
to a volume in $k$-space equal to $2 \pi k \times k_{x,{\rm max}}
\times \Delta k$, i.e.\ $\delta P \propto k^{-1/2}$.  This contrasts
with a fully spectroscopic survey, for which the relevant Fourier
modes for a scale $k$ reside within a spherical shell, such that
$\delta P \propto k^{-1}$.  Although a photometric redshift survey
maps out a reduced volume of Fourier space, the larger
density-of-states (owing to the increased cosmic volume probed) can
still result in a more accurate measurement of the galaxy power
spectrum.

In order to illustrate this point, we compare the power spectrum
accuracies for our simulated photometric redshift surveys with those
expected for the SDSS spectroscopic surveys (both the main survey and
the Luminous Red Galaxy (LRG) survey).  We created Monte Carlo power
spectrum realizations for these SDSS spectroscopic surveys using the
methodology of Blake \& Glazebrook (2003), which is very similar to
that presented in Section \ref{secmeth} above.  The main differences
are that no photometric redshift smearing is applied, and thus a
conical geometry may be employed rather than a flat-sky approximation
(although as discussed in Section \ref{secapprox}, this makes a
negligible difference to the results). We modelled the SDSS main
spectroscopic survey using the redshift distribution
\begin{equation}
\frac{dN}{dz} \propto z^2 \exp{ \left[ - \left( \frac{z}{0.055}
\right)^{1.31} \right] }
\end{equation}
with a total surface density equal to $\Sigma_0 = 70.7$
deg$^{-2}$. This model constitutes a good fit to the relevant
luminosity function (Blanton et al.\ 2003).  The redshift interval of
the simulation was $0 < z < 0.25$. We approximated the LRG
spectroscopic survey redshift distribution using the Gaussian function
\begin{equation}
\frac{dN}{dz} \propto \exp{ \left[ - \left( \frac{z - 0.375}{0.065}
\right)^2 \right] }
\end{equation}
and a total surface density $\Sigma_0 = 17.2$ deg$^{-2}$, which
provides a reasonable fit to the radial selection function discussed
by Eisenstein et al.\ (2001).  The redshift interval of the simulation
was $0.3 < z < 0.45$ and LRGs are assigned a linear bias factor $b=2$
(see Equation \ref{eqpk}). For both spectroscopic surveys we assumed
an areal coverage of $10{,}000$ deg$^2$, the same as for the simulated
photometric redshift surveys.

As illustrated by Figure \ref{figpksurv}, in the turnover regime ($k <
0.02 \, h$ Mpc$^{-1}$) the photometric redshift surveys always yield
more large-scale modes than the SDSS spectroscopic surveys owing to
the larger cosmic volume mapped and the fact that the wavelengths of
these modes significantly exceed the length-scale of
photometric-redshift radial smearing.  In the acoustic oscillations
regime ($k > 0.05 \, h$ Mpc$^{-1}$), a $10{,}000$ deg$^2$ photometric
redshift survey out-performs the SDSS LRG spectroscopic survey
provided that $\sigma_0 \la 0.03$ and $r_{\rm lim} \ga 22.5$.

\section{Measuring the acoustic oscillations using photometric redshift
surveys}
\label{secwig}

\begin{figure*}
\center
\epsfig{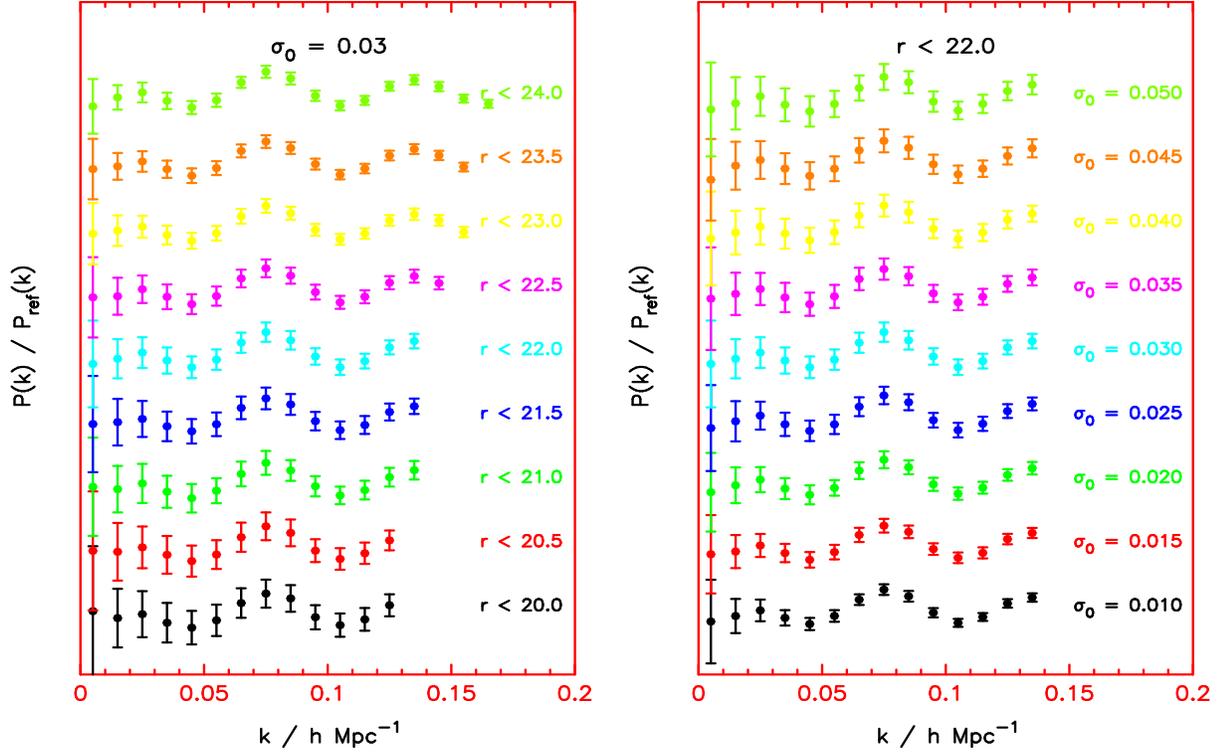}
\caption{Mean values of the power spectrum divided by the smooth
  reference spectrum, $P(k)/P_{\rm ref}$, averaged over the Monte
  Carlo realizations for various photometric redshift surveys,
  together with the standard deviation in each bin.  The left-hand
  panel illustrates the variation of the observed $P(k)/P_{\rm ref}$
  as a function of $r_{\rm lim}$ for $\sigma_0 = 0.03$; the right-hand
  panel displays the dependence on $\sigma_0$ for $r_{\rm lim} = 22$.
  Results for different surveys are offset for clarity.  The power
  spectra are plotted for wavelengths larger than the
  linear/non-linear transition scale $k_{\rm lin}$.  This scale varies
  with the threshold magnitude $r_{\rm lim}$ because it is evaluated
  at the effective redshift of the survey $z_{\rm eff} = (z_{\rm min}
  + z_{\rm max})/2$ (see Table \ref{tabsurv}).  In all cases we use a
  survey area of $10{,}000$ deg$^2$.}
\label{figpkwig}
\end{figure*}

The clustering power spectrum on intermediate scales ($0.05 < k < 0.3
\, h$ Mpc$^{-1}$) contains small-amplitude ($\sim 5$ per cent)
modulations known as `acoustic oscillations' (Peebles \& Yu 1970; Hu
\& Sugiyama 1996; Eisenstein \& Hu 1998).  These fluctuations in power
have an identical physical origin to those observed in the CMB,
namely, oscillations in the photon-baryon fluid before recombination.

There has been considerable recent interest in exploiting these
acoustic features as an accurate and clean probe of the cosmological
model (Blake \& Glazebrook 2003; Seo \& Eisenstein 2003; Linder 2003;
Hu \& Haiman 2003).  The approximately sinusoidal fluctuations in
power encode a characteristic scale -- {\it the sound horizon at
  recombination} -- which can be measured from the CMB.  This scale
can then act as a {\it standard cosmological ruler} (Eisenstein, Hu \&
Tegmark 1999): its recovered value from a galaxy redshift survey
depends on the assumed cosmological parameters, in particular the dark
energy model, and may be used to constrain those parameters in a
manner that is probably significantly less sensitive to systematic
error than other probes (Blake \& Glazebrook 2003).

Very recently, analysis of the clustering pattern of SDSS Luminous Red
Galaxies at $z \approx 0.35$ has yielded the first convincing
detection of the acoustic signal and application of the standard ruler
(Eisenstein et al.\ 2005).  Although this survey does not have
sufficient redshift reach to constrain dark energy models, this result
is an important validation of the technique.  Indeed, {\it detection
  of these acoustic features represents a fundamental test of the
  paradigm of the origin of galaxies in the fluctuations observed in
  the CMB.}

Utilization of the acoustic oscillations to measure the properties of
dark energy demands new galaxy surveys of unprecedented depth and
volume (Blake \& Glazebrook 2003; Seo \& Eisenstein 2003).  Given the
current availability of large-scale imaging surveys such as the SDSS,
and the anticipated wait of several years for commencement of projects
with sufficiently capable spectroscopic facilities able to survey
$\sim 10^6$ objects over $\sim 1000$ deg$^2$ (such as the KAOS
proposal), it is timely to evaluate the role photometric redshift
surveys could play in the detection and measurement of the acoustic
oscillations.  Furthermore, recently-developed novel
photometric-redshift techniques such as those utilizing {\it
  artificial neural networks} should prove extremely useful in this
regard (e.g.\ Collister \& Lahav 2004).

The constraints on the cosmological model yielded by acoustic
oscillations in future photometric redshift surveys have been
discussed by Seo \& Eisenstein (2003); Amendola et al.\ (2004) and
Dolney et al.\ (2004). Here we take a different but complementary
approach.  Firstly, this previous work deduced cosmological parameter
constraints using a Fisher matrix approach which provides the {\it
  minimum possible} errors for an unbiased estimate of a given
parameter, based upon the curvature of the likelihood surface near the
fiducial model.  In the present study we instead use Monte Carlo
techniques, which make a closer connection with the analysis methods
that would be used for real data and can probe more realistic
non-parabolic likelihood surfaces. Secondly, we give detailed
consideration to the statistical confidence of detection of the
relevant power spectrum features, carefully separating this
information from that contained in the overall shape of the power
spectrum, which may be subject to additional systematic distortions,
as discussed below. Thirdly, by treating a wide grid of potential
photometric redshift surveys varying both the redshift accuracy and
the limiting magnitude, we can make a direct connection with the
performance of current and future experiments.

The comparison of photometric redshift and spectroscopic redshift
surveys has already been discussed in detail by Blake \& Glazebrook
(2003) and Seo \& Eisenstein (2003).  To summarize the relevant points
of these two papers:
\begin{itemize}
\item As discussed in Section \ref{secpksteps}, Fourier modes with
  values of $k_x \gg 1/\sigma_x$ (where $x$ is the radial axis)
  contribute noise.  A photometric redshift survey therefore requires
  significantly more sky area than a spectroscopic redshift survey of
  similar depth to yield the same number of Fourier modes in a given
  power spectrum bin with scale $k \gg 1/\sigma_x$.
\item Fourier modes with usable signal-to-noise ratios are largely
  tangential ($k_x \approx 0$).  Consequently, in the case of a
  photometric redshift survey, we are only able to apply the standard
  ruler represented by the acoustic oscillations in the tangential
  direction, constraining the co-ordinate distance $x(z)$ to the
  effective redshift of the survey.  We lose the capacity of a
  spectroscopic redshift survey to apply the ruler radially, measuring
  $dx/dz$ (or equivalently the Hubble constant at redshift $z$), which
  yields powerful additional constraints on the dark energy model.
\end{itemize}

In this Section we present a series of simulations addressing the
issues of the {\it confidence} and {\it accuracy} of detection of
acoustic oscillations as a function of photometric redshift error
$\sigma_0$ (as defined by Equation \ref{eqzerr}) and limiting apparent
magnitude $r_{\rm lim}$ of the imaging survey.  We defer the questions
of whether and how these requirements can be realized in realistic
surveys to Section \ref{secrealsurv}.  We proceed in a
model-independent fashion, quantifying the statistical significance
with which we can assert deviations from a featureless monotonic
function, without using any of the information contained in the power
spectrum shape.  In Section \ref{seccospar} we use a more
model-dependent approach, combining the full power spectrum shape
function with recent measurements of the CMB anisotropies to derive
predicted constraints on the cosmological parameters.

Our set of Monte-Carlo realized power spectra enables us to evaluate
statistical questions of confidence and accuracy over a realistic
ensemble of Universes, without needing to approximate the statistical
distributions or likelihood surfaces, except that when converting
values of the $\chi^2$ statistic to relative probabilities we
implicitly assume that the errors in the measured power spectra are
Gaussian, which agrees well with histograms obtained from the Monte
Carlo realizations.

\subsection{Confidence of detection of acoustic oscillations}

We note that {\it confidence of detection} can be defined in several
different ways and depends strongly on statistical priors.  One
approach to the data analysis would be to fit full $\Lambda$CDM
transfer functions (e.g.\ the formulae of Eisenstein \& Hu 1998) to
the measured power spectra and thereby determine that baryonic models
(containing acoustic features) provided a significantly better fit to
the data than models with $\Omega_{\rm b} = 0$.  We argue in the
current Section that this only partially constitutes a detection of
acoustic oscillations, because information contained in the {\it
  shape} of the power spectrum is also constraining this fit.

We adopted a conservative approach in which, prior to measuring the
preferred sinusoidal scale, we divided the measured power spectra by a
smooth `wiggle-free' reference spectrum.  For our purposes, this is
the `no-wiggles' spectrum of Eisenstein \& Hu (1998) (see Blake \&
Glazebrook (2003); with real data, additional smooth polynomial terms
can be fitted to remove any residual shape).  We therefore do not
utilize any information encoded by the shape of the power spectrum.
The purpose of our philosophy is to maintain maximum independence from
models and systematic effects: the shape of $P(k)$ may be subject to
smooth broad-band systematic tilts induced by such effects as complex
biasing schemes, a running primordial spectral index, and
redshift-space distortions.  For the acoustic oscillations analysis,
the power spectrum is measured in bins of width $\Delta k = 0.01 \, h$
Mpc$^{-1}$.  Plots of simulated power spectra divided by reference
spectra for different survey configurations are displayed in Figure
\ref{figpkwig}.

The resulting sinusoidal modulation for each realization is fitted
with a simple 2-parameter empirical formula describing a decaying
sinusoid, i.e.\ Equation 3 from Blake \& Glazebrook 2003:
\begin{equation}
\frac{P(k)}{P_{\rm ref}} = 1 + A \, k \, \exp{\left[ - \left(
\frac{k}{0.1 \, h \, {\rm Mpc}^{-1}} \right)^{1.4} \right]} \,
\sin{\left( \frac{2 \pi k}{k_A} \right)}.
\label{eqwigfit}
\end{equation}
For each realization, we recorded (i) the best-fitting characteristic
scale $k_A$, (ii) the value of the chi-squared statistic for the
`no-wiggles' model (i.e. Equation \ref{eqwigfit} with $A = 0$),
$\chi^2_{\rm no-wig}$, and (iii) the value of the chi-squared
statistic for the best-fitting `wiggles' model, $\chi^2_{\rm
  wig-best}$.  The $\chi^2$ statistic was defined in the usual manner:
\begin{equation}
\chi^2 = \sum_i \left( \frac{P_{\rm obs}(k_i) - P_{\rm
model}(k_i)}{\sigma_P(k_i)} \right) ^2 .
\end{equation}
Our flat-sky approximation implies that the off-diagonal terms of the
covariance matrix (i.e.\ correlations between adjacent power spectrum
bins) are consistent with zero; this was explicitly tested by
computing full covariance matrices for a test case.

\begin{figure}
\center
\epsfig{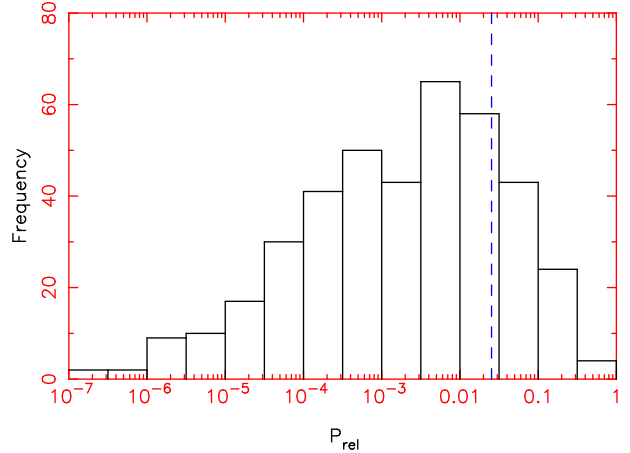}
\caption{Histogram of values of $P_{\rm rel}$ (defined by Equation
  \ref{eqprelwig}) for 400 Monte Carlo realizations of a survey with
  $r_{\rm lim} = 21.5$ and $\sigma_0 = 0.03$, illustrating the skewed
  distribution of probabilities (note the logarithmic $x$-axis).  The
  mean value of $P_{\rm rel}$ (as plotted in Figure
  \ref{figwigconfphoto}) is indicated by the vertical dashed line;
  realizations possess less confident detections of acoustic
  oscillations than implied by this mean in just 19 per cent of
  cases.}
\label{figprobdist}
\end{figure}

A simple relative probability of the `no-wiggles' model and `wiggles'
model can be defined by
\begin{equation}
P_{\rm rel} = \exp{ [ - (\chi^2_{\rm no-wig} - \chi^2_{\rm
wig-best})/2 ]}
\label{eqprelwig}
\end{equation}
(but see below for further discussion).  We note that the distribution
of values of $P_{\rm rel}$ across Monte Carlo realizations of the
Universe is far from symmetric, as illustrated by Figure
\ref{figprobdist}.  In a Universe falling at the 50th percentile of
the ensemble, the relative probability of the `no-wiggles' model
compared to the `wiggles' model would be significantly lower than the
mean of the distribution.

\begin{figure}
\center
\epsfig{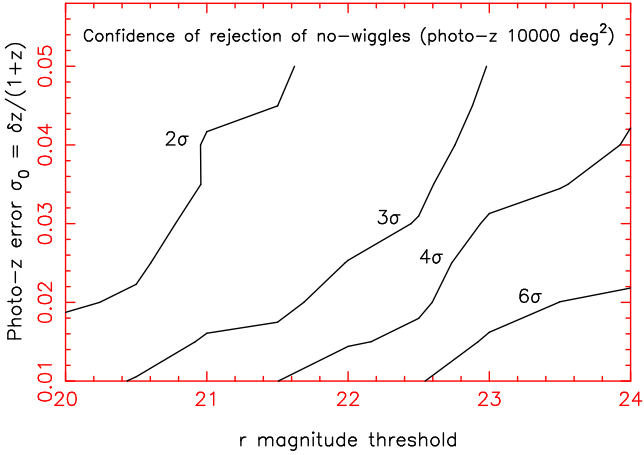}
\caption{Contours of the detection confidence of acoustic oscillations
  defined by the average value of $P_{\rm no-wig} = P_{\rm rel}/(1 +
  P_{\rm rel})$ (Equation \ref{eqprelwig}) for photometric-redshift
  imaging surveys with varying magnitude threshold $r_{\rm lim}$ and
  photometric-redshift error parameter $\sigma_0$.  The probabilities
  are expressed as a rejection `number of sigmas' for a Gaussian
  distribution.  We use a survey area of $10{,}000$ deg$^2$.}
\label{figwigconfphoto}
\end{figure}

As our initial assessment of the confidence of detection of the
acoustic oscillations we considered the average value of the quantity
$P_{\rm rel}$ defined by Equation \ref{eqprelwig} over the Monte Carlo
realizations.  We converted this into a probability for the
`no-wiggles' model by using $P_{\rm rel} = P_{\rm no-wig}/P_{\rm wig}$
and $P_{\rm no-wig} + P_{\rm wig} = 1$.  Contours of $P_{\rm no-wig}$
(expressed as a rejection `number of sigmas' for a Gaussian
distribution) are displayed in Figure \ref{figwigconfphoto} in the
parameter space of $(r_{\rm lim},\sigma_0)$.  In order to obtain a
3-$\sigma$ detection confidence of $99.7$ per cent ($P_{\rm no-wig} =
3 \times 10^{-3}$) we require a survey with parameters such that
\begin{equation}
\sigma_0 \la (r_{\rm lim} - 19.5) \times 0.01 .
\end{equation}

\begin{figure}
\center
\epsfig{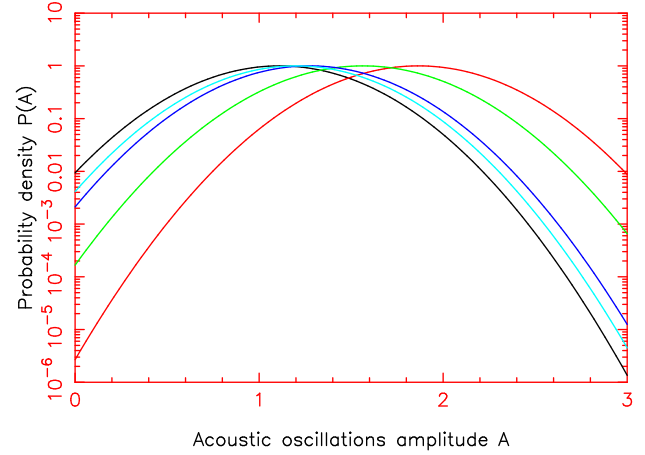}
\caption{Probability distributions (as defined by Equation
  \ref{eqprobamp}) of amplitude $A$ for the first 5 Monte Carlo
  realizations of a survey with parameters $r_{\rm lim} = 21.5$ and
  $\sigma_0 = 0.03$.  The curves are normalized such that $P(A_{\rm
    best}) = 1$.  The range of intercepts at $A = 0$ illustrates the
  distribution of values for $P_{\rm no-wig}$ across the Monte Carlo
  realizations.}
\label{figprobamp}
\end{figure}

As an alternative method of quantifying the `confidence of detection'
of acoustic oscillations (i.e.\ the probability that $A \ne 0$ in
Equation \ref{eqwigfit}) we considered the following Bayesian
approach. We placed a uniform prior Prior(A) on the value of $A$:
\begin{eqnarray}
{\rm Prior}(A) &=& \frac{1}{A_{\rm wid}} \,\,\,\,\,\,\,\, A_{\rm min}
< A < A_{\rm min} + A_{\rm wid} \nonumber \\ &=& 0 \,\,\,\,\,\,\,\,
{\rm elsewhere}.
\end{eqnarray}
We chose $A_{\rm min} = 0$ and $A_{\rm wid} = 3 \approx 2 A_{\Lambda
  {\rm CDM}}$, to be conservative.  We assumed prior knowledge of the
acoustic wavescale $k_A = 2\pi/s$, where $s$ is the value of the sound
horizon at recombination, known very accurately from linear CMB
physics (e.g.\ Eisenstein \& Hu 1998, Equation 26).  For an individual
power spectrum realization, the probability density as a function of
amplitude $A$ is
\begin{equation}
P(A) \propto \exp{(-\chi^2/2)}
\label{eqprobamp}
\end{equation}
where $\chi^2$ is the value of the chi-squared statistic of the fit of
Equation \ref{eqwigfit} to the data of that realization (with $k_A =
2\pi/s$).  Figure \ref{figprobamp} displays curves of $P(A)$ against
$A$ for the first few Monte Carlo realizations of a simulated survey
with $r_{\rm lim} = 21.5$ and $\sigma_0 = 0.03$.

\begin{figure}
\center
\epsfig{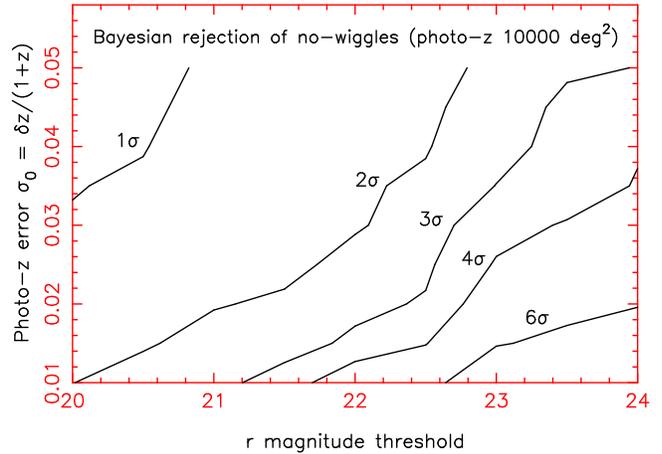}
\caption{Contours of the detection confidence of acoustic oscillations
  defined by the average value of the Bayesian quantity $P_{\rm
    no-wig}$ (Equation \ref{eqprelbayes}) for photometric-redshift
  imaging surveys with varying magnitude threshold $r_{\rm lim}$ and
  photometric-redshift error parameter $\sigma_0$.  The probabilities
  are expressed as a rejection `number of sigmas' for a Gaussian
  distribution.  We use a survey area of $10{,}000$ deg$^2$.}
\label{figwigconfbayes}
\end{figure}

According to Bayesian statistics, the relative probability of a
`no-wiggles' and `wiggles' model for one realization is:
\begin{equation}
\frac{P_{\rm no-wig}}{P_{\rm wig}} = \frac{P(A=0)} { \left(
  \int_{-\infty}^{\infty} P(A) \, {\rm Prior}(A) \, dA \right)}
\label{eqprelbayes}
\end{equation}
In the numerator of this expression, a $\delta$-function prior centred
at $A = 0$ has been integrated over.  Figure \ref{figwigconfbayes}
plots the average value over the Monte Carlo realizations of the
quantity $P_{\rm no-wig}$ defined by Equation \ref{eqprelbayes}
(expressed as a rejection `number of sigmas' for a Gaussian
distribution) in the parameter space of $(r_{\rm lim},\sigma_0)$.
Note that {\it less confident} detections of the acoustic oscillations
are implied by using this prior, requiring surveys approximately half
a magnitude deeper for a 3-$\sigma$ detection.  This is reasonable
because the probability density of the `non-detection' model with $A =
0$ is being compared to the average probability density of models with
$A \ne 0$, rather than only to the best-fitting `detection' model.
This serves to illustrate the critical role of priors in quantifying
the `confidence of detection'.  Note that if we had widened our prior
yet further by allowing a range in possible model acoustic oscillation
scales $k_A$ then an even higher-performance survey would be required
to achieve the same detection confidence.  An alternative prior on the
amplitude for the `wiggles' model would have been to use a delta
function ${\rm Prior}(A)=\delta(A-A_{\Lambda {\rm CDM}})$, which would
produce very similar results to using $P_{\rm rel}$ (of equation
\ref{eqprelwig}).

\subsection{Accuracy of measurement of acoustic oscillations}

\begin{figure}
\center
\epsfig{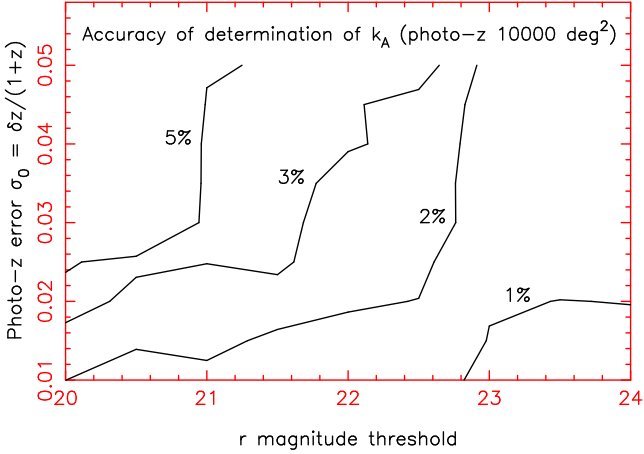}
\caption{Contours of the accuracy of determination of the
  characteristic scale $k_A$ for photometric-redshift imaging surveys
  with varying magnitude threshold $r_{\rm lim}$ and
  photometric-redshift error parameter $\sigma_0$.  We use a survey
  area of $10{,}000$ deg$^2$.}
\label{figwigaccphoto}
\end{figure}

We can also use our simulations to quantify the accuracy with which
the characteristic scale (i.e.\ standard ruler) can be recovered from
a photometric redshift survey, as a function of $r_{\rm lim}$ and
$\sigma_0$.  This is easily obtained as the spread in best-fitting
values of $k_A$ over the Monte Carlo realizations.  We defined this
spread as half the difference between the 16th and 84th percentiles of
the distribution of fitted wavelengths.  This quantity is plotted in
Figure \ref{figwigaccphoto} as a percentage fractional precision
$\Delta k_A/k_A$ in the parameter space of $(r_{\rm lim},\sigma_0)$.
The precision improves with both increasing $r_{\rm lim}$ and
decreasing $\sigma_0$, peaking at $\approx 0.7$ per cent for our
highest-performance survey $(r_{\rm lim} = 24, \sigma_0 = 0.01)$.

In cosmological terms, this precision is equal to the accuracy with
which the quantity $x(z_{\rm eff})/s$ can be determined by the survey
(where $x$ is the co-ordinate distance to the effective redshift of
the survey and $s$ is the value of the sound horizon at
recombination).  This may in turn be converted into confidence
distributions for dark energy models (e.g.\ Seo \& Eisenstein 2003;
Amendola et al.\ 2004; Dolney et al.\ 2004; Glazebrook \& Blake 2005).

\subsection{Comparison with spectroscopic surveys}

It is of great interest to compare the confidence and accuracy of the
acoustic oscillation measurement from putative photometric redshift
surveys with those resulting from future spectroscopic redshift
surveys.  We therefore created Monte Carlo power spectrum realizations
of a grid of spectroscopic surveys, using the same techniques as our
photometric survey analysis.  We varied the total survey area
$A_\Omega$ (from $1000$ deg$^2$ to $10{,}000$ deg$^2$) and the
limiting magnitude threshold $r_{\rm lim}$ (from $18$ to $24$).

For our spectroscopic survey analyses we assumed the same redshift
distributions as a function of $r_{\rm lim}$ listed in Table
\ref{tabsurv}, although we note that a realistic spectroscopic survey
would more likely be directed at a sub-population such as star-forming
galaxies with strong emission lines, which would be selected in a more
complex manner than a simple magnitude cut.

\begin{figure}
\center
\epsfig{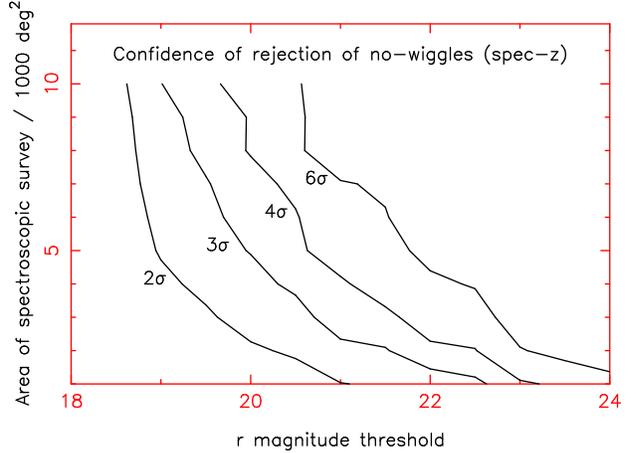}
\caption{Contours of the detection confidence of acoustic oscillations
  defined by the average value of $P_{\rm no-wig} = P_{\rm rel}/(1 +
  P_{\rm rel})$ (Equation \ref{eqprelwig}) for spectroscopic redshift
  surveys with varying magnitude threshold $r_{\rm lim}$ and survey
  area $A_\Omega$.  The probabilties are expressed as a rejection
  `number of sigmas' for a Gaussian distribution.  This plot may be
  compared directly with Figure \ref {figwigconfphoto} for photometric
  redshift surveys (note the different ranges of $x$-axis).}
\label{figwigconfspec}
\end{figure}

Our spectroscopic redshift power spectra were analyzed for acoustic
oscillation measurement in an identical manner to the photometric
redshift surveys.  For purposes of comparison we bin power spectra
averaging over angles, and do not separate the results into tangential
and radial components.  Figure \ref{figwigconfspec} displays the
confidence of detection as a function of $(A_\Omega,r_{\rm lim})$,
quantified by the value of $P_{\rm no-wig}$ in the same manner as
Figure \ref{figwigconfphoto}.  A 3-$\sigma$ detection of the acoustic
oscillations can be achieved by a spectroscopic survey with parameters
($A_\Omega = 1000$ deg$^2$, $r_{\rm lim} = 22.5$) or ($A_\Omega =
3000$ deg$^2$, $r_{\rm lim} = 21$).  For comparison, an equivalent
detection is yielded by a $10{,}000$ deg$^2$ photometric redshift
survey with parameters ($\sigma_0 = 0.05$, $r_{\rm lim} = 23$) or
($\sigma_0 = 0.01$, $r_{\rm lim} = 20.5$).  Note that the confidences
of detection listed here are more conservative (by up to a factor of
four in terms of the number of standard deviations for a Gaussian
distribution) than those which would result from a full fit of a
$\Lambda$CDM model power spectrum, as discussed and compared in
Section \ref{secbarfrac}.

\begin{figure}
\center
\epsfig{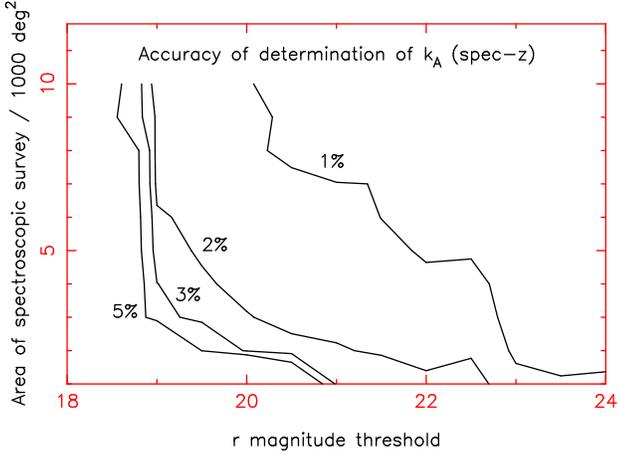}
\caption{Contours of the accuracy of determination of the
  characteristic scale $k_A$ for spectroscopic redshift surveys with
  varying magnitude threshold $r_{\rm lim}$ and survey area
  $A_\Omega$.  This plot may be compared directly with Figure \ref
  {figwigaccphoto} for photometric redshift surveys (note the
  different ranges of $x$-axis).}
\label{figwigaccspec}
\end{figure}

Figure \ref{figwigaccspec} displays the resulting accuracy of
measurement of the characteristic acoustic scale; this plot may be
compared directly with Figure \ref{figwigaccphoto}.  For example, a
spectroscopic survey of depth $r_{\rm lim} \approx 22.5$ over
$A_\Omega \approx 1000$ deg$^2$ will achieve a $2\%$ measurement of
the standard ruler (a similar precision is achieved by a $10{,}000$
deg$^2$ photo-z survey with the same depth and redshift error
$\sigma_0 = 0.03$).

\begin{figure}
\center
\epsfig{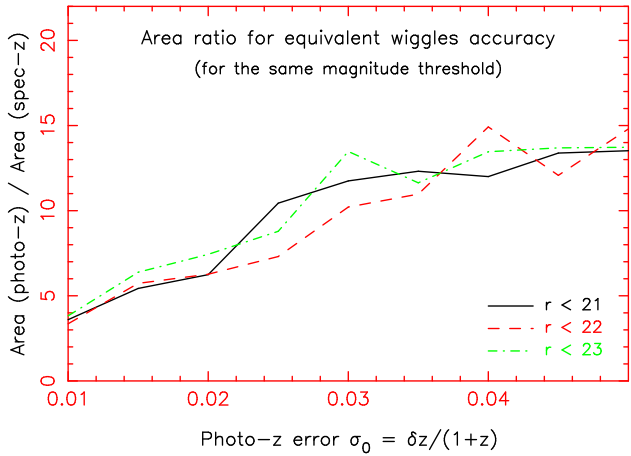}
\caption{The area ratio of photometric and spectroscopic redshift
  surveys achieving the same accuracy of standard ruler measurement
  for a fixed magnitude threshold.  This factor is determined by the
  photometric redshift precision.}
\label{figspecvsphot1}
\end{figure}

\begin{figure}
\center
\epsfig{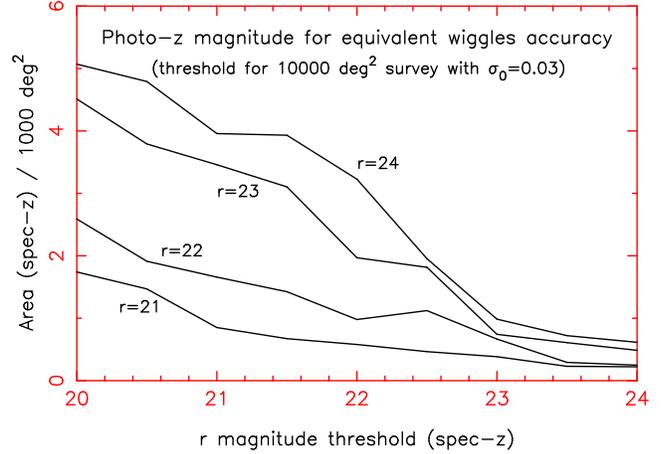}
\caption{The magnitude threshold required by a $10{,}000$ deg$^2$
  photometric redshift survey (with $\sigma_0 = 0.03$) to match the
  standard ruler accuracy of a grid of spectroscopic surveys with
  varying area and depth.}
\label{figspecvsphot2}
\end{figure}

Figures \ref{figspecvsphot1} and \ref{figspecvsphot2} continue the
comparison of photometric and spectroscopic surveys.  In Figure
\ref{figspecvsphot1} we plot the ratio of areas of photometric and
spectroscopic surveys achieving the same accuracy of standard ruler
measurement {\it for a fixed magnitude threshold common to both
  surveys}.  We assume an area of $10{,}000$ square degrees for the
photometric redshift survey and vary the area of the spectroscopic
redshift survey, although the results are expected to apply more
generally.  From Figure \ref{figspecvsphot1} we see that, for a
photometric redshift precision $\sigma_0 = 0.03$, the area ratio for a
fixed magnitude threshold is about a factor of 12.  This is simply
understood by the requirement that the number of Fourier modes
contributing to the power spectrum measurement ($m \propto k_{x,{\rm
    max}} A_\Omega \propto A_\Omega/\sigma_0$) must be roughly equal
in the two cases.  For example, if $\sigma_0 = 0.03$ then $k_{x,{\rm
    max}} \approx 0.02 \, h$ Mpc$^{-1}$.  However, for a spectroscopic
survey $k_{x,{\rm max}} = k_{\rm lin} \approx 0.2 \, h$ Mpc$^{-1}$,
thus the same number of modes $m$ is delivered by a survey area
$A_\Omega$ reduced by a factor of $\approx 10$.  The relation $m
\propto A_\Omega/\sigma_0 \approx \rm{constant}$ also explains the
overall scaling $A_\Omega \propto \sigma_0$ apparent in Figure
\ref{figspecvsphot1}.

A comparison at common magnitude threshold is of course simplistic:
for given observational resources, an imaging survey can readily probe
to fainter magnitudes.  Therefore, Figure \ref{figspecvsphot2}
considers a grid of spectroscopic surveys (parameterized by
$A_\Omega$, $r_{\rm lim}$) and illustrates the magnitude depth
required by a $10{,}000$ deg$^2$ photometric survey with $\sigma_0 =
0.03$ to match the standard ruler accuracy.  The contours ($r_{\rm
  photo} = 21 \rightarrow 24$) correspond to standard ruler accuracies
in the range $5 \rightarrow 1.5$ per cent (see Figure
\ref{figwigaccphoto}).

\section{Measuring the turnover using photometric redshift surveys}
\label{secturn}

\begin{figure*}
\center
\epsfig{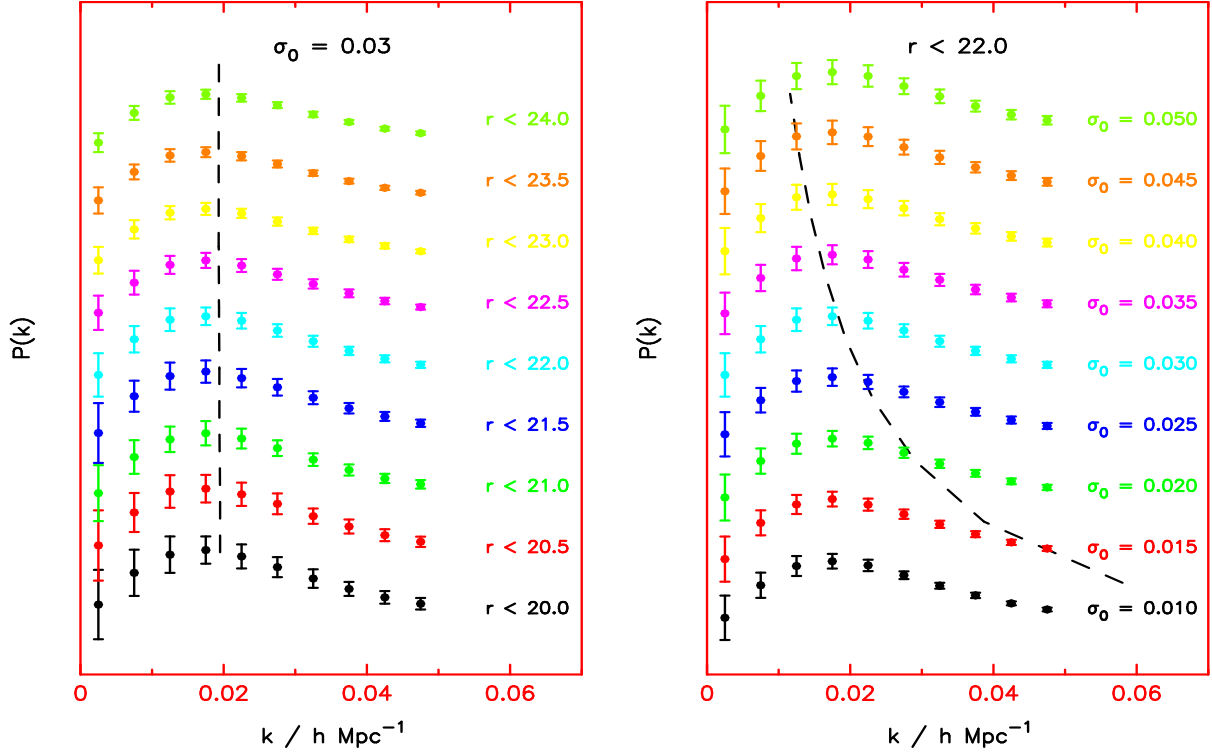}
\caption{Mean values of the power spectrum $P(k)$ averaged over the
  Monte Carlo realizations for various surveys, together with the
  standard deviation in each bin.  In this plot we do not divide by
  any smooth reference power spectrum.  The left-hand panel
  illustrates the variation of the obtained $P(k)$ as a function of
  $r_{\rm lim}$ for $\sigma_0 = 0.03$; the right-hand panel displays
  the dependence on $\sigma_0$ for $r_{\rm lim} = 22$.  Results for
  different surveys are offset for clarity.  The dashed line tracks
  the value of $k_{x,{\rm max}}$ for the surveys in question.  For a
  bin at $k < k_{x,{\rm max}}$, no further improvement in power
  spectrum precision is possible via more accurate values of
  $\sigma_0$, rather a survey must go deeper, mapping more cosmic
  volume and increasing the density of states in $k$-space.  Power
  spectra are only plotted for scales $k < 0.05 \, h$ Mpc$^{-1}$.}
\label{figpkturn}
\end{figure*}

According to standard cosmological theory the clustering power
spectrum should exhibit a `turnover' (i.e.\ a maximum in power) at a
characteristic scale $k_{\rm turn} \approx 0.015 \, h$ Mpc$^{-1}$. The
turnover arises because the primordial power spectrum laid down by
cosmological inflation -- hypothesized to be a featureless power law
$P_{\rm prim}(k) \propto k^{n_{\rm scalar}}$, where $n_{\rm scalar}
\approx 1$ -- is suppressed by a `transfer function' $T(k)$ owing to
radiation pressure in the radiation-dominated epoch.  The resulting
linear-theory power spectrum is derived as $P(k) \propto P_{\rm
  prim}(k) \, T(k)^2$ where $T(k) \approx 1$ for $k < k_{\rm turn}$
and $T(k)$ decreases towards zero for $k > k_{\rm turn}$ with an
approximate limiting dependence $k^{-2}$.  The characteristic scale
$k_{\rm turn}$ is equivalent to the {\it co-moving horizon scale at
  matter-radiation equality}.  This is sensitive to the quantity
$\Omega_{\rm m} h^2$, since the larger the physical density of matter
($\Omega_{\rm m} h^2$), the earlier matter-radiation equality occurs,
and suppression of growth due to radiation oscillations below the
Jeans length does not have time to reach larger scales.  Therefore the
scale of the turnover is smaller and $k_{\rm turn}$ is larger.  (Note
that since we measure redshift and not distance, the $x-$axis of the
power spectrum plot is in units of $h$ Mpc$^{-1}$ and so the position
of the turnover on a plot of $P(k)$ against $k/h$ depends on
$\Omega_{\rm m} h$).  Moreover, structure modes with wavelengths
larger than the turnover scale are relatively unaffected by physics
subsequent to inflation and potentially constitute a probe of the
inflationary epoch.

Detection of the turnover in the galaxy clustering pattern constitutes
an interesting test of the cosmological paradigm.  Its absence may
imply either a failure of the standard cosmological theory, or the
discovery of new large-scale galaxy biasing mechanisms (e.g.\ Durrer
et al.\ 2003; i.e., the turnover is more susceptible to systematic
distortions than the acoustic oscillations). Successful definition of
the turnover requires a survey possessing an extremely large volume to
reduce the effect of cosmic variance.  In addition, the number density
of the tracer galaxies must be sufficient to suppress the shot noise
contribution to the power spectrum error. For example, quasi-stellar
objects can easily be detected to high redshift but possess an
inadequate number density to permit an experiment limited by cosmic
variance (Outram et al.\ 2003).  As a result, no survey has cleanly
measured the turnover yet (in a model-independent manner).

{\it For a turnover detection experiment, spectroscopic-redshift
  accuracy is not required.}  The relevant scales are sufficiently
large that equivalent information may be recovered from a photometric
redshift survey, if the main contribution to the photometric redshift
errors is statistical and not systematic.  For example, for a survey
with $\sigma_0 = 0.03$ and $z_{\rm eff} = 0.5$, all Fourier modes with
$k_x < k_{x,{\rm max}} \approx 0.02 \, h$ Mpc$^{-1} > k_{\rm turn}$
survive the radial damping.

We now consider the detectability of the turnover and implied accuracy
of determination of the characteristic scale $k_{\rm turn}$ for a
series of simulations varying $r_{\rm lim}$ and $\sigma_0$.  As
before, our starting point is the ensemble of power spectrum
realizations obtained as described in Section \ref{secpksteps}.  For
the turnover analysis, the power spectrum was measured in bins of
width $\Delta k = 0.005 \, h$ Mpc$^{-1}$.  Because the turnover in
$P(k)$ occurs at an approximate scale of $k_{\rm turn} \approx 0.015
\, h$ Mpc$^{-1}$, we only utilize power spectrum modes with $k < 0.04
\, h$ Mpc$^{-1}$ (i.e.\ 8 bins).  Plots of simulated power spectra for
different survey configurations are displayed in Figure
\ref{figpkturn}.

As with the acoustic oscillations analysis, the significance of
detection of the turnover depends on our prior assumptions.  We again
take a conservative approach, fitting our realized power spectra with
a simple empirical parabolic turnover model characterized by four
parameters:
\begin{eqnarray}
P(k) &=& P_0 \left[ 1 - \alpha \left( \frac{k - k_0}{k_0} \right)^2
\right] \hspace{1cm} (k < k_0) \nonumber \\ &=& P_0 \left[ 1 - \beta
\left( \frac{k - k_0}{k_0} \right)^2 \right] \hspace{1cm} (k > k_0)
\label{eqturnfit}
\end{eqnarray}
The free parameters are the turnover scale $k_0$, the maximum of the
power spectrum $P_0$, and the amplitudes of the parabolic decrease of
power on either side of the maximum, $\alpha$ and $\beta$.  In this
sense, a detection of the turnover is determined by finding a
best-fitting value for $\alpha$ significantly greater than zero, and
is governed solely by power spectrum modes at scales larger than the
turnover scale (i.e.\ $k < k_0$).  The requirement that $P(k) \ge 0$
restricts the fitted parameters to lie in the ranges $P_0 \ge 0$,
$\alpha \le 1$, $\beta \le 1$ and we implement these conditions as
strong priors in our fitting process.

\begin{figure}
\center
\epsfig{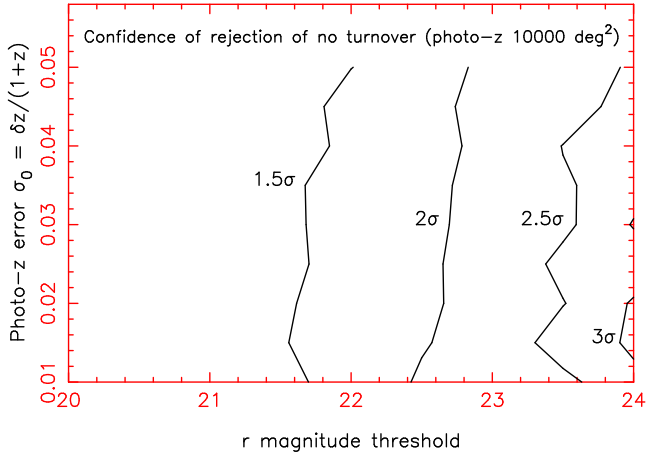}
\caption{Contours of turnover detection confidence defined by the
  average value of $P_{\rm no-turn} = P_{\rm rel}/(1 + P_{\rm rel})$
  (Equation \ref{eqprelturn}) for photometric-redshift imaging surveys
  with varying magnitude threshold $r_{\rm lim}$ and
  photometric-redshift error parameter $\sigma_0$.  The probabilities
  are expressed as a rejection `number of sigmas' for a Gaussian
  distribution.  We use a survey area of $10{,}000$ deg$^2$.}
\label{figturnconfphoto}
\end{figure}

By analogy with the acoustic oscillations analysis, we defined the
confidence of turnover detection for a given power spectrum
realization by comparing the chi-squared statistic for the
best-fitting turnover model (i.e.\ Equation \ref{eqturnfit}) with that
for the best-fitting `no turnover' model, which we defined as Equation
\ref{eqturnfit} with $\alpha$ set equal to zero:
\begin{equation}
P_{\rm rel} = \exp{ [ - (\chi^2_{\rm no-turn} - \chi^2_{\rm
turn-best})/2 ]}
\label{eqprelturn}
\end{equation}
Figure \ref{figturnconfphoto} plots the average value of $P_{\rm
  no-turn} = P_{\rm rel} / (1 + P_{\rm rel})$ over the Monte Carlo
realizations (expressed as a rejection `number of sigmas' for a
Gaussian distribution) in the parameter space of $(r_{\rm
  lim},\sigma_0)$.  In our highest-performance survey ($r_{\rm lim} =
24$, $\sigma_0 = 0.01$), the turnover is detected with $\approx 99.5$
per cent confidence ($P_{\rm no-turn} = 0.005$).  A 2-$\sigma$
detection in the mean realization requires $r_{\rm lim} \approx 22.5$.
We note that, provided $\sigma_0 \la 0.04$, the detection confidence
is independent of the photometric redshift accuracy because all
Fourier modes beyond the turnover are retained in our analysis.

\begin{figure}
\center
\epsfig{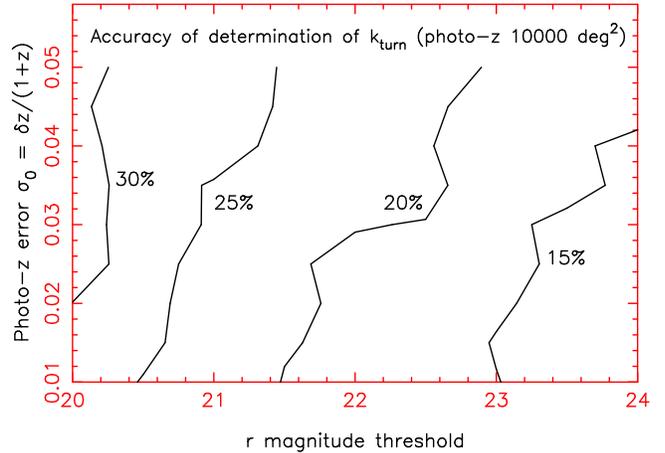}
\caption{Contours of the accuracy of determination of the
  characteristic turnover scale $k_{\rm turn}$ for
  photometric-redshift imaging surveys with varying magnitude
  threshold $r_{\rm lim}$ and photometric-redshift error parameter
  $\sigma_0$.  We use a survey area of $10{,}000$ deg$^2$.}
\label{figturnaccphoto}
\end{figure}

Figure \ref{figturnaccphoto} quantifies the accuracy with which the
characteristic turnover scale $k_{\rm turn}$ can be recovered from a
photometric redshift survey, as a function of $r_{\rm lim}$ and
$\sigma_0$, using the same technique as for the characteristic
acoustic oscillation scale in Section \ref{secwig}.  In the best case
we considered, the turnover scale can be measured with a precision of
$\approx 12$ per cent.  This is considerably poorer than the
measurement accuracy of the acoustic oscillations scale, owing to the
broadness of the turnover and the vastly fewer Fourier modes available
at the relevant scales.  This observation could in principle yield a
$12\%$ measurement of $\Omega_{\rm m} h$.  This in itself is not
particularly competitive with other techniques, but performing the
analysis in this model-independent way tests the fundamental
assumptions made in the standard analyses, and isolates the possible
influences of relevant systematic effects.  For example, it may
indicate that scale-dependent biasing occurs on large scales
(e.g.\ Dekel \& Rees 1987) which would constitute a critical
observation in the field of galaxy formation.  Furthermore, precise
measurements of the large-scale clustering pattern may unveil
currently-unknown signatures of inflation or of non-Gaussianity
(e.g.\ Martin \& Ringeval 2004) pointing to a new cosmological
paradigm.

Unlike the case of acoustic oscillations measurement, we do not make a
comparison with spectroscopic redshift surveys for our turnover
detection experiment, because all relevant Fourier modes are retained
by a photometric redshift survey and therefore the results would be
unchanged if perfect redshifts were known.  Thus photometric redshift
surveys will always be more efficient for turnover detection if the
systematic errors can be controlled.

\section{Measurements of the cosmological parameters}
\label{seccospar}

In this Section we adopt a more model-dependent analysis approach to
that performed in Sections \ref{secwig} and \ref{secturn}, assuming
the full theoretical framework of $\Lambda$CDM transfer functions and
calculating how our simulated measurements of the galaxy power
spectrum from photometric redshift surveys can be used to constrain
cosmological parameters more tightly.  Our investigation here is thus
independent, but complementary, to the results presented earlier, and
indicates how the cosmological conclusions are tightened by the
incorporation of more model assumptions.

We assume linear biasing, i.e.\ that the galaxy power spectrum is a
constant multiple of the matter power spectrum, and marginalize over
this parameter with a flat prior. We assume that the bias parameter
does not evolve with redshift, and discuss the effect of this
approximation in Section~\ref{secapprox}.

In order to search parameter space we use Markov-Chain Monte Carlo
(MCMC) sampling using the Metropolis-Hastings algorithm.  In short, an
MCMC `chain' is made up of a list of `samples' (coordinates in
parameter space) which are obtained from performing trial likelihood
evaluations ${\rm Pr}({\bf x})$.  A new sample at position ${\bf
  x}_{i+1}$ is accepted with a probability ${\rm min} ({\rm Pr}({\bf
  x}_{i+1}) /{\rm Pr}({\bf x}_i),1)$.  The difficulty lies in
suggesting good trial positions; we use the latest version of CosmoMC
(see Lewis \& Bridle 2002 and {\tt http://cosmologist.info/cosmomc}
for more information) which uses CAMB (Lewis, Challinor \& Lasenby
2000) to calculate CMB and matter power spectra.  We ran MCMC chains
for the least powerful photometric redshift survey and used importance
sampling to find parameter constraints for the better surveys.

We used flat priors on the CMB parameters $\Omega_{\rm b} h^2$,
$\Omega_{\rm c} h^2$, $\theta_{\rm peak}$, $\tau$, $n_{\rm s}$,
$n_{\rm run}$ and $\log 10 (A_{\rm s})$.  $\theta_{\rm peak}$ is used
instead of the Hubble constant $h$ because it renders the MCMC method
more efficient when CMB data is included; it is defined by approximate
formulae which given the CMB first peak position in terms of the other
cosmological parameters.  Therefore for a given set of cosmological
parameters, $\theta_{\rm peak}$ can be converted into $h$, and vice
versa.  For the simulations in this Section we take as our fiducial
parameters those from the abstract of Spergel et al.\ (2003): $h =
0.72$, $\Omega_{\rm b} h^2 = 0.024$, $\Omega_{\rm m} h^2 = 0.14$,
$\tau = 0.16$, $n_{\rm s} = 0.99$.  The widths of the priors are
chosen to be sufficiently large that they have no influence on the
results.  We assume adiabatic initial conditions with a negligible
tensor contribution.

\subsection{Measurements of the baryon fraction}
\label{secbarfrac}

\begin{figure}
\center
\epsfig{file=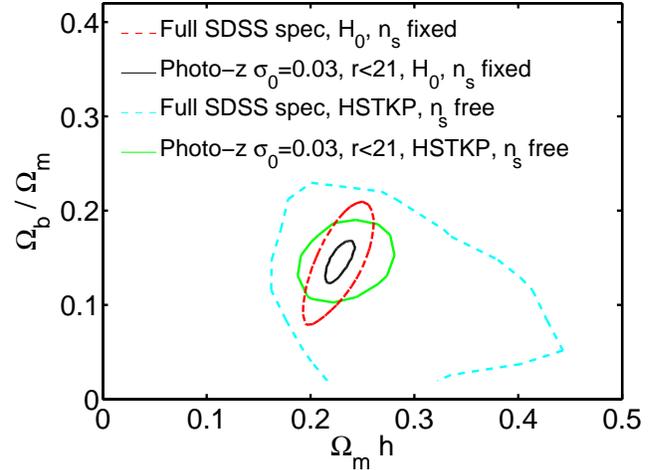,width=8.5cm,angle=0}
\caption{Contours in the parameter space of ($\Omega_{\rm
    b}/\Omega_{\rm m}$, $\Omega_{\rm m} h$) derived from galaxy power
  spectrum measurements alone.  To generate the red and black (darker)
  lines we fixed other cosmological parameters at their input values.
  The dashed red (darker) contour is for our projected SDSS
  spectroscopic survey and the inner solid black (darker) contour is
  for a photometric survey with $\sigma_0=0.03$ and $r<21$. For the
  cyan and green (lighter) lines $H_0$ and $n_{\rm s}$ are
  marginalized over using an HST Key Project prior of $72 \pm 8$ km
  s$^{-1}$ Mpc$^{-1}$.  Only 68\% confidence contours are shown, for
  clarity.}
\label{figobom_omh}
\end{figure}

First we considered constraints on the baryon fraction resulting from
galaxy surveys alone, as a function of $\Omega_{\rm m} h$.  We fixed
all other cosmological parameters at their input values for ease of
comparison with current galaxy power spectrum analyses (e.g.\ Cole et
al.\ 2005).  We derived results for simulated power spectra from the
SDSS main spectroscopic survey (using the survey parameters listed in
Section \ref{secpkerr}) and for an SDSS photometric redshift imaging
survey (assuming $\sigma_0 = 0.03$, $r_{\rm lim} = 21$) using sky area
$10{,}000$ deg$^2$ in both cases.  These contours are plotted as the
darker lines in Figure~\ref{figobom_omh}.  Photometric redshifts from
the SDSS imaging survey would produce tighter parameter constraints
than the final SDSS spectroscopic survey, if redshift accuracy
$\sigma_0 = 0.03$ to a magnitude limit of $r_{\rm lim} = 21$ could be
achieved.  This is a challenging requirement, but may be approachable
by selecting Luminous Red Galaxies.

In order to quantify the contribution of the overall shape to the
detection of $\Omega_{\rm b}/\Omega_{\rm m}$ we added in as free
parameters the Hubble constant and the primordial power spectrum tilt
parameter $n_{\rm s}$.  We applied an HSTKP prior on the Hubble
constant, otherwise the Hubble constant had significant probability
above 100 km s$^{-1}$ Mpc$^{-1}$.  The new constraints from the
projected SDSS spectroscopic and photometric redshift surveys are
shown by the lighter contours in Figure~\ref{figobom_omh}.  This is a
more rigorous test of detection of acoustic oscillations -- hence the
slightly wider contours.  The effect of relaxing the assumptions on
$n_{\rm s}$ and $H_0$ is small when the constraints are weak, since
the prior that $\Omega_{\rm b}>0$ already limits the maximum error
bar.

We quantify the constraint on the baryon density by calculating the
probability as a function of $\Omega_{\rm b}/\Omega_{\rm m}$
marginalized over $\Omega_{\rm m} h$ and any other parameters.  We
then find the error bar by halving the distance between equi-probable
limits containing 68 per cent of the probability and quote a `number
of sigma' by dividing the fiducial value of $\Omega_{\rm
  b}/\Omega_{\rm m}$ by this error bar.  The `number of sigma' for our
full SDSS spectroscopic survey simulation decreases from 3.5 to 2.3 on
allowing $H_0$ to vary within the HSTKP prior and completely freeing
the spectral index $n_{\rm s}$.  For the example photometric redshift
survey shown here, it changes from 9 to 5.

We note that the constraints on the baryon density from our analysis
are rather weak from the main SDSS survey, despite the fact that it is
larger than the full 2dFGRS survey for which the baryon density has
already been detected with better confidence.  This is because we have
assumed a relatively conservative value for the maximum wavenumber
fitted by the linear power spectrum, $k_{\rm max} = 0.11 \, h$
Mpc$^{-1}$ (see Section \ref{secpksteps}).  The exact results are
quite sensitive to this value.  In practice, experimental teams may
choose to use a larger value of $k_{\rm max}$, obtaining tighter
contours than those displayed in Figure \ref{figobom_omh}, but
increasing their sensitivity to the systematic uncertainties of
modelling the quasi-linear regime.  For comparison with existing
measurements (e.g.\ Cole et al.\ 2005) and for maximum contrast with
the `model independent' sections, we fix $H_0$ and $n_{\rm s}$ for the
remainder of this Section.

\begin{figure}
\center
\epsfig{file=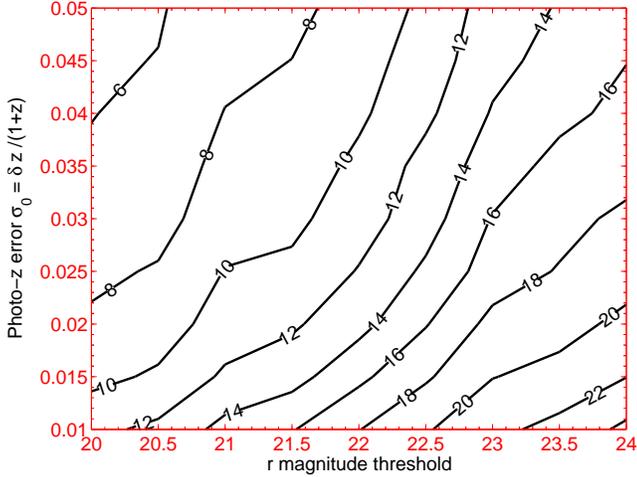,width=8.5cm,angle=0}
\caption{Input value of $\Omega_{\rm b}/\Omega_{\rm m}$ divided by the
  68 per cent confidence error margin for $\Omega_{\rm b}/\Omega_{\rm
    m}$ (loosely speaking, the `number of $\sigma$ of detection') for
  a range of photometric redshift surveys with varying magnitude
  threshold $r$ and photometric-redshift error parameter $\sigma_0$.
  The shape parameter $\Omega_m h$ is varied, as for the darker
  contours in Figure \ref{figobom_omh}.  We use a survey area of
  $10{,}000$ deg$^2$.}
\label{figconf_obom}
\end{figure}

Figure~\ref{figconf_obom} displays how the detection confidence of
$\Omega_{\rm b}/\Omega_{\rm m}$ depends on the parameters of a general
set of future photometric redshift surveys.  We conclude that the
significance of measurement of a non-zero value for $\Omega_{\rm
  b}/\Omega_{\rm m}$ will shortly be greatly improved by the use of
photometric redshifts.  For a detection with 6-$\sigma$ confidence we
require a survey with $r_{\rm lim} \approx 20$ and $\sigma_0 \la
0.04$.

\begin{figure}
\center
\epsfig{file=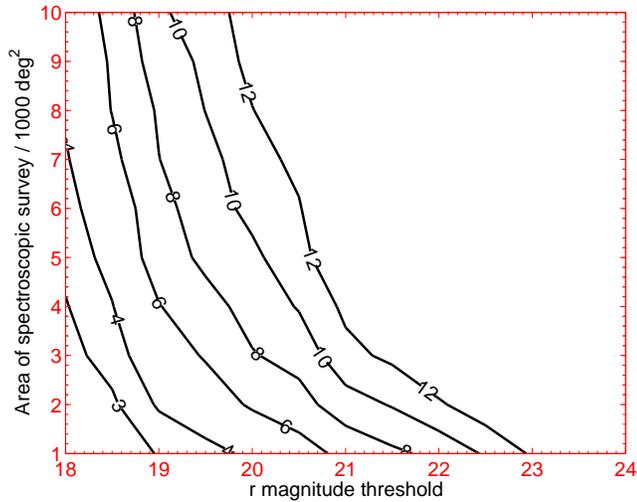,width=8.5cm,angle=0}
\caption{Input value of $\Omega_{\rm b}/\Omega_{\rm m}$ divided by the
  68 per cent confidence error margin for $\Omega_{\rm b}/\Omega_{\rm
    m}$ (loosely speaking, the `number of $\sigma$ of detection') for
  a range of spectroscopic redshift surveys with varying magnitude
  threshold $r$ and area.  The shape parameter $\Omega_m h$ is varied,
  as for the darker contours in Figure \ref{figobom_omh}.  }
\label{figconf_obom_spec}
\end{figure}

\begin{figure}
\center
\epsfig{file=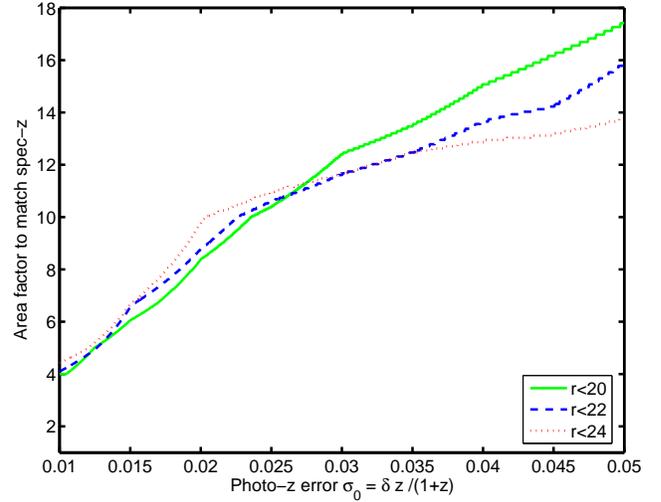,width=8.5cm,angle=0}
\caption{Area factor required to obtain the same accuracy of
  measurement of $\Omega_{\rm b}/\Omega_{\rm m}$ for a photometric
  redshift survey as for a spectroscopic survey.}
\label{figconf_obom_speccfphot}
\end{figure}

Equivalent confidences are shown in Figure \ref{figconf_obom_spec} for
a general set of future deeper spectroscopic redshift surveys.
Clearly for a given magnitude limit, the area required to achieve a
detection of $\Omega_{\rm b}/\Omega_{\rm m} \ne 0$ is smaller than the
$10{,}000$ deg$^2$ used for the photometric survey simulations with
the same magnitude limit, however to survey this area is significantly
more costly.  A quantitative comparison between the results for
spectroscopic and photometric redshift surveys is shown in Figure
\ref{figconf_obom_speccfphot}.  This is derived from the previous two
figures by calculating the area of spectroscopic survey required to
obtain the same detection confidence as the $10{,}000$ square degree
imaging survey, for each magnitude limit and photometric redshift
error.  From this figure it can be seen that the factor in area
required to make up for the photometric redshift uncertainties is
about $12 (\sigma_0/0.03)$ and roughly independent of magnitude limit,
in good agreement with the `model independent' analysis.

The value of $\Omega_{\rm b}/\Omega_{\rm m}$ is connected to the
presence of the acoustic oscillations in the matter power spectrum,
and thus, as discussed in Section \ref{secwig}, a high-significance
measurement of a non-zero value for $\Omega_{\rm b}/\Omega_{\rm m}$
could be considered as a potential `detection' of the existence of
these acoustic features.  Our treatment in the current Section is thus
more model-dependent than and complementary to our analysis method of
Section~\ref{secwig}.  In the present Section we implicitly assume as
a prior the entire $\Lambda$CDM framework in which the acoustic
oscillation position, amplitude and matter power spectrum shape are
intrinsically linked; whereas in Section~\ref{secwig} we adopted a
more conservative approach, simply fitting a modified sinusoidal
function to the simulated data.

Therefore it is not surprising that our detection confidences for
$\Omega_{\rm b}/\Omega_{\rm m} \ne 0$ are somewhat tighter than those
for a `model-independent' detection of acoustic oscillations.
Roughly, the 8-$\sigma$ lines in Figure~\ref{figconf_obom} lie on the
2-$\sigma$ lines of Figure~\ref{figwigconfphoto}.  Generally the
number of $\sigma$ is a factor of four larger for the
`model-dependent' fit with $h$ and $n_{\rm s}$ held fixed.  As
illustrated by Figure \ref{figobom_omh}, if we instead marginalize
over the Hubble constant, with an HSTKP prior and free $n_{\rm s}$,
then the measurements are much less accurate.  The $r < 21$, $\sigma_0
= 0.03$ survey constraint weakens from 9-$\sigma$ to 5-$\sigma$ on
freeing $h$ and $n_{\rm s}$ in this way, compared to $2.2$-$\sigma$
for the fully `model-independent' fit.

The two different analysis methods presented in this study
(i.e.\ Sections \ref{secwig} and \ref{seccospar}) are analogous to
those used to detect the CMB EE polarization signal (DASI, Kovac et
al.\ 2002; CBI, Readhead et al.\ 2004).  For the first detection of a
non-zero signal, the DASI team used a `template' for the shape of the
EE power spectrum, taken from the $\Lambda$CDM model that best fit the
CMB TT data.  By contrast, the CBI team were the first to detect the
phase and amplitude of the EE polarization signal using a
model-independent sine-wave fit, and found that the inferred values
were consistent with a $\Lambda$CDM model.  In order to test the
framework of the $\Lambda$CDM model we argue that a precise
measurement of baryon wiggles in the matter power spectrum using a
model-independent fit will be an enormous break-through; this being
done, more assumptions can then be made to extract the most accurate
possible measurements of the cosmological parameters.

\subsection{Measurements of the running spectral index}

A vital role for measurements of the galaxy power spectrum is to break
the parameter degeneracies inherent in the Cosmic Microwave Background
(CMB) anisotropies. Using a standard six-parameter $\Lambda$CDM model,
the WMAP satellite measurements can be readily converted to
constraints on the matter power spectrum (e.g.\ Tegmark 2003), which
are sometimes erroneously interpreted to mean that there is no need
for galaxy redshift surveys.  However, the most important current
questions in cosmology, such as the quest to quantify the properties
of the dark energy and of inflation, demand that the simplest
cosmological model be extended to encompass additional parameters,
such as a time-varying dark energy equation-of-state and a more
general model for the primordial power spectrum of mass fluctuations.
In these cases the degeneracies inherent in the CMB become
insuperable, and high-quality additional data is essential.

In the standard six-parameter cosmological model, it is assumed that
the scalar perturbations have a power-law power spectrum parameterized
by a single spectral index $n_{\rm s}$.  However, the simplest
inflationary models predict that this index should exhibit a slight
dependence on scale, often parameterized $n_{\rm run}$, such that the
primordial power spectrum assumes the form
\begin{equation}
P(k) \propto (k/k_0)^{n_{\rm s} + n_{\rm run} \log{(k/k_0)}}
\label{eqnrun}
\end{equation}
where according to standard inflationary models, $n_{\rm run} \sim
0.002$. This parameter $n_{\rm run}$ has been the subject of much
recent debate due to the apparent detection of a non-zero value by the
WMAP team (Spergel et al.\ 2003) at $n_{\rm run} = -0.031 \pm 0.024$
from WMAP, 2dFGRS, ACBAR and CBI. Therefore we include $n_{\rm run}$
as an additional parameter in our analysis, expanding the total number
of fitted parameters to seven (noting that the presence of a running
spectral index is potentially degenerate with that of a
scale-dependent bias).

\begin{figure}
\center
\epsfig{file=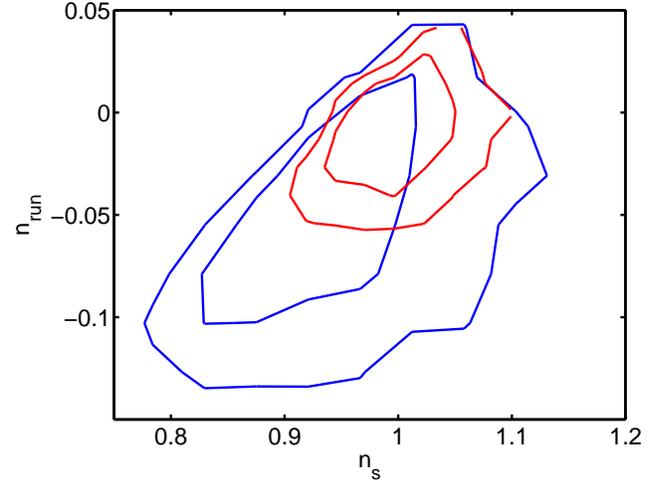,width=8.5cm,angle=0}
\caption{Confidence contours in the parameter space of ($n_{\rm run}$,
  $n_{\rm s}$) resulting from our simulated SDSS spectroscopic survey
  (outer contours) and from a simulated SDSS photometric redshift
  survey with parameters $r_{\rm lim} = 21$ and $\sigma_0 = 0.03$
  (inner contours).  In each case, the galaxy power spectrum data is
  combined with WMAP CMB data and we marginalized over the other five
  cosmological parameters.}
\label{fignsnrun}
\end{figure}

Figure~\ref{fignsnrun} plots the resulting constraints on the
primordial power spectrum parameters of Equation \ref{eqnrun} --
$n_{\rm s}$ and $n_{\rm run}$ -- when the WMAP CMB data is combined
with our simulated $10{,}000$ deg$^2$ SDSS main spectroscopic and
photometric redshift ($\sigma_0 = 0.03$) surveys.  The more accurate
power spectrum measurements yielded by the photometric redshift survey
helps break the degeneracy between the scalar spectral index $n_{\rm
  s}$ and the running spectral index $n_{\rm run}$ (although the exact
direction of the degeneracy is determined by the pivot scale in
Equation~\ref{eqnrun}, for which we use $k_0 = 0.05$ Mpc$^{-1}$).

\begin{figure}
\center
\epsfig{file=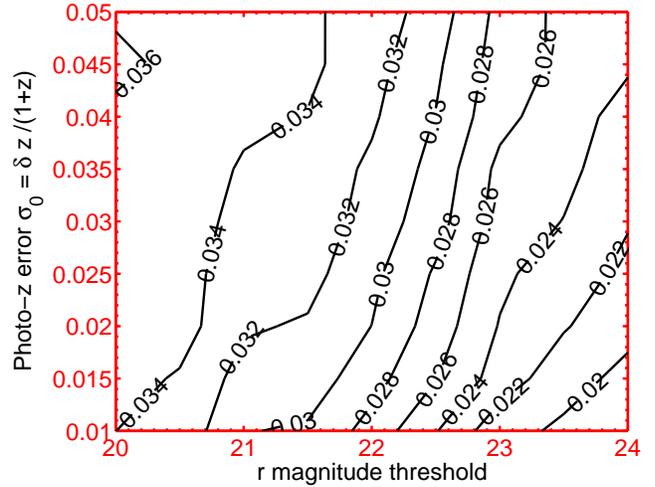,width=8.5cm,angle=0}
\caption{68 per cent confidence error margin of $n_{\rm run}$ for a
  range of photometric redshift surveys with varying magnitude
  threshold $r$ and photometric-redshift error parameter $\sigma_0$.
  In all cases we combined with WMAP data and marginalized over the
  remaining cosmological parameters $H_0$, $\Omega_{\rm b} h^2$,
  $\Omega_{\rm m} h^2$, $\tau$, $\sigma_8$ and $n_{\rm s}$.}
\label{fignrun}
\end{figure}

Figure~\ref{fignrun} displays the (1-$\sigma$) accuracy of measurement
of $n_{\rm run}$ for a general set of future photometric redshift
surveys. For comparison, our projection for the full SDSS
spectroscopic sample is $\pm 0.037$.  The constraint would be slightly
tighter if more CMB data were included in the analysis.  The error in
$n_{\rm run}$ is halved for our best possible imaging survey case
($r_{\rm lim} = 24$, $\sigma_0 = 0.01$), bringing the limit in between
that predicted by slow-roll inflation and that indicated by the WMAP
first-year results.

\begin{figure}
\center
\epsfig{file=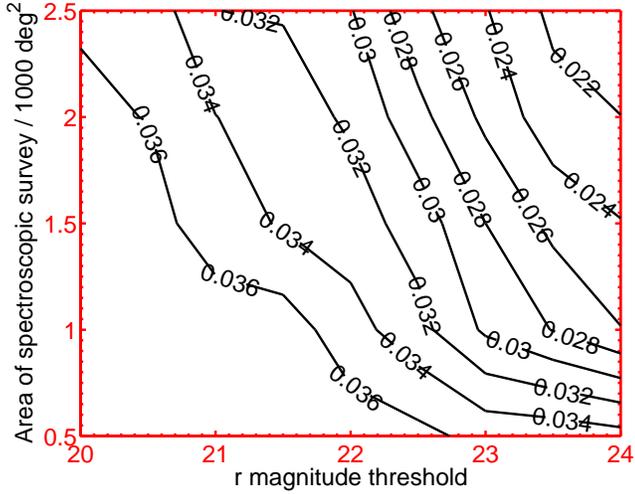,width=8.5cm,angle=0}
\caption{68 per cent confidence error margin of $n_{\rm run}$ for a
  range of spectroscopic redshift surveys with varying magnitude
  threshold $r$ and area.  In all cases we combined with WMAP data and
  marginalized over the remaining cosmological parameters $H_0$,
  $\Omega_{\rm b} h^2$, $\Omega_{\rm m} h^2$, $\tau$, $\sigma_8$ and
  $n_{\rm s}$, as in Figure~\ref{fignrun}.}
\label{fignrun_spec}
\end{figure}

In Figure~\ref{fignrun_spec} we show the equivalent constraints for a
general set of future spectroscopic redshift surveys.  We see that to
achieve an error bar $\delta n_{\rm run} = 0.03$, a spectroscopic
survey requires about 15 per cent of the area of a photometric
redshift survey with $\sigma_0=0.03$ to the same magnitude limit.

\begin{figure}
\center
\epsfig{file=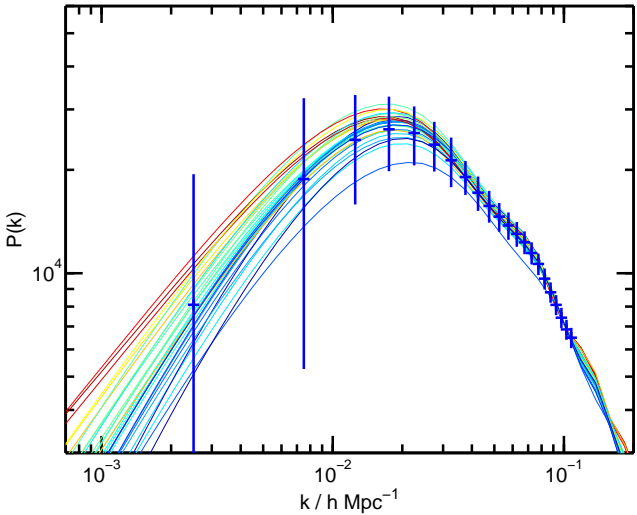,width=8.5cm,angle=0}
\caption{The range of model matter power spectra allowed by the WMAP
  data combined with a photometric redshift survey with parameters
  $\sigma_0 = 0.03$ and $r<21$, colour-coded according to the value of
  $n_{\rm run}$.  The plotted data points display the simulated
  photometric redshift power spectrum.  Seven cosmological parameters
  are allowed to vary: $H_0$, $\Omega_{\rm b} h^2$, $\Omega_{\rm m}
  h^2$, $\tau$, $\sigma_8$, $n_{\rm s}$ and $n_{\rm run}$.  Since we
  marginalize over the linear bias parameter then for display purposes
  the power spectra are normalized to go through the highest $k$ data
  point.} \label{figpknrun}
\end{figure}

As discussed above, detection of the matter power spectrum shape on
the largest scales is important because it has been unchanged since
inflation.  In Figure~\ref{figpknrun} we indicate the range of model
matter power spectra permitted by the WMAP data (Verde et al. 2003,
Hinshaw et al. 2003, Kogut et al. 2003) plus our fiducial SDSS
photometric redshift survey, for a seven-parameter $\Lambda$CDM model
(i.e.\ including a free parameter $n_{\rm run}$).  This shows that
despite the improvements in power spectrum precision, there is still
some freedom in our knowledge of the matter power spectrum on large
scales.

Further relaxations of the post-inflation assumptions will heighten
the importance of matter power spectrum information. Whilst the
increasing amount of CMB polarization information will help to improve
constraints, the range of possible models may be widened even
further. For example: in the above we have assumed that the
perturbations are adiabatic, with a negligible tensor
contribution. Moreover, in addition to adding tensors, a number of
isocurvature modes are possible, along with freedom in their spatial
correlations (e.g.\ Bucher et al.\ 2004).  In addition each mode has a
power spectrum that could be relaxed from the power law and gentle
running forms assumed here.  We would ideally like to reconstruct
these power spectra, or equivalently the inflationary potentials. For
example a phase transition during inflation can cause a step-like
feature in the scalar power spectrum, and trans-Planckian effects can
induce `ringing'. This additional freedom can be constrained
effectively by combining CMB and large-scale structure information
(e.g.\ Mukherjee \& Wang 2003; Bridle et al.\ 2003).

\section{Assessing our approximations}
\label{secapprox}

We now assess the effect of the most significant approximations
contained in our methodology for simulating the accuracy of power
spectrum measurements (Section \ref{secmeth}).

\subsection{Photometric redshift error distribution}
\label{secphotozerr}

Our fiducial set of simulations assumed that the statistical
distribution of photometric redshift errors could be described by a
Gaussian function characterized by a standard deviation (Equation
\ref{eqzerr}).  This spread can always be measured by obtaining
spectra of a complete sub-sample of the imaged galaxies.

However, a real flux-limited survey will inevitably contain a
combination of different classes of galaxy with different intrinsic
photometric-redshift scatters.  For example, Luminous Red Galaxies
(LRGs) have especially strong spectral breaks which yield improved
photometric redshift precision compared to an average galaxy
possessing the same redshift and $r$-band magnitude.

We assessed the effect of such combinations via several test cases in
which the photometric redshift error distribution was modelled by a
sum of two Gaussian functions with different widths (denoted
$\sigma_1$, $\sigma_2$) and relative amplitudes (denoted $b_1$, $b_2$
such that $b_1 + b_2 = 1$).  Specifically, we assumed $\sigma_1 =
0.01$, $\sigma_2 = (0.03,0.1)$ and $b_2 = (0.1,0.25)$, such that we
are investigating the effect of a {\it minority sub-population with a
  significantly broader error distribution than the majority of
  galaxies}.

Our motivation is to verify that the single Gaussian error function in
Section \ref{secmeth} does not yield over-optimistic results relative
to a more realistic double-Gaussian model with the same standard
deviation, i.e.\ to check that our analysis is conservative.  Thus in
each case we compared the fractional power spectrum precision
resulting from the double Gaussian model with that of a single
Gaussian error distribution with the same overall standard deviation
($\sigma_{\rm eff} = \sqrt{b_1 \sigma_1^2 + b_2 \sigma_2^2}$).

For the double Gaussian model we derived the value of $k_{x,{\rm
    max}}$ in Equation \ref{eqkxmax} by taking the value of $\sigma_x$
corresponding to the tighter of the two Gaussians (i.e.\ the dominant
galaxy population).  We determined by experiment that this was optimal
compared to other possible choices, such as $\sigma_x$ corresponding
to the overall standard deviation $\sigma_{\rm eff}$.

We found that the precision of power spectrum measurement was never
degraded by the assumption of a double-Gaussian model, and in the case
($\sigma_2 = 0.1$, $b_2 = 0.1$) was significantly improved compared to
our fiducial predictions for a single-Gaussian model owing to the
tighter core of the photometric redshift error distribution.  In other
words our previous analysis is indeed conservative: the presence in
real survey data of a minority sub-population of galaxies with
significantly poorer photometric redshift precision than the overall
standard deviation will not result in poorer power spectrum
measurements than those inferred by the equivalent single Gaussian
function.

We also experimented with the addition of an overall systematic offset
$z_{\rm off}$ to the mean photometric redshift (relative to the
spectroscopic value) of the second of the two Gaussian functions.  For
the case ($\sigma_2 = 0.03$, $b_2 = 0.1$) the power spectrum precision
was degraded by $\approx 10\%$ when $z_{\rm off} = 0.05$.  We note
that such systematic effects can always be identified by spectroscopic
follow-up of a large enough complete sub-sample of the imaging survey.

\subsection{Bias model}

Inference of the clustering pattern from galaxy surveys is always
subject to uncertainties associated with the bias model, i.e.\ the
precise manner in which galaxy light traces the underlying mass
fluctuations.  In our initial simulations we assumed that this biasing
scheme was simply linear; but in general the bias mechanism will be
non-linear, scale-dependent, non-local and evolving with redshift.

Observationally, linear scale-independent bias appears a good
approximation for a wide range of galaxy types on the large scales
discussed here (e.g.\ Peacock \& Dodds 1994): structure formation is
still in the linear regime and the physics of individual galaxy
formation should not be relevant.  We note however that halo-dependent
effects may become important for the most massive galaxies (Peacock \&
Smith 2000; Seljak 2000).

Even if bias is scale-dependent on large scales, it would be very
surprising if it induced oscillatory features in $k$-space liable to
obscure the distinctive acoustic peaks and troughs.  Our
model-independent analysis of acoustic oscillations detection in
Section \ref{secwig} should be robust to such systematic broadband
tilts in the galaxy power spectrum because the overall shape is
divided out.  Some authors have argued that scale-dependent bias on
large scales should obscure the turnover (Durrer et al.\ 2003); if
this property was confirmed observationally then it would undermine
our turnover detection analysis of Section \ref{secturn}, but would
tell us something interesting about galaxy formation.

In addition, we assume that the biasing scheme does not depend on
redshift.  This is in conflict with observational data: Lahav et
al.\ (2002) assume a model in which the clustering of galaxies remains
constant with time (in co-moving space) whilst the dark matter
perturbations grow, implying $b(z) = b_0/D(z)$ where $D(z)$ is the
linear growth factor.  Alternatively, if galaxies are assumed to
follow the cosmic flow and remain constant in number then an
alternative functional form is produced (Fry 1996) in which the linear
bias parameter evolves in proportion to its deviation from unity in
the present-day universe according to $b(z)=1+(b_0-1)/D(z)$.  Our
fiducial model is $b(z) = 1$, which is a special case of the Fry
equation, and in fitting to the data we have assumed the rather
simplistic form $b(z) = b_0$.

We note that despite their differences, all of these models have the
same number of free parameters, namely one, $b_0$.  Therefore if we
were to fit instead the Fry bias model to a simulation with $b(z) = 1$
then we would have the same amount of information available for
extracting cosmological parameters.  Clearly $b(z) = 1$ is not a
solution of the `constant galaxy clustering' (CGC) model of Lahav et
al.\ (2002), but relative to this model our simulation and detections
are conservative, since if $b_0 \sim 1$ today (as measured for the
2dFGRS by Lahav et al.\ 2002 and Verde et al.\ 2002) and $b(z) =
b_0/D(z)$ (CGC model) then the galaxy clustering at higher redshift
would be stronger than in our simulation and thus the signal to noise
available for acoustic oscillation and turnover detection would be
increased.

We also point out that: \\ $\bullet$ For a power spectrum measurement
limited by cosmic variance (such as those discussed here), the
fractional error in the power spectrum (e.g.\ as plotted in Figure
\ref{figpkwig}) is independent of the overall amplitude of the power
spectrum as determined by the bias model. \\ $\bullet$ An incorrect
assumption about the bias model would not shift the acoustic
oscillations along the wavenumber axis; the features would remain in
the same place and add up coherently when integrating over redshift.

\subsection{Curved-sky effects}

Our flat-sky approximation would not be a valid analysis method for a
real $10{,}000$ deg$^2$ imaging survey: we must either break such a
survey into smaller sky areas, or utilize a spherical harmonics
decomposition along the lines of Percival et al.\ (2004).  However,
given that the fractional power spectrum accuracy of our simulated
surveys is determined principally by the cosmic volume surveyed, our
results should be representative.  In order to verify this, we
re-analyzed the SDSS main spectroscopic survey discussed in Section
\ref{secpkerr} by casting the volume in a flat-sky box rather than a
conical geometry (this case is at very low redshift, hence is
particularly sensitive to this flat-sky approximation).  The resulting
power spectrum precision agreed to within 10 per cent.

\subsection{Analysis in redshift slices}

Application of the flat-sky analysis method to a real photometric
redshift survey (with the redshift errors assumed to smear the
distribution along one axis only) must additionally involve the survey
being split into redshift slices: any significant ($\ga 20\%$)
variation of the angular diameter distance across an analyzed box
would cause a smoothing-out of the acoustic features.  Again our
results should be representative because, regardless of the number of
redshift slices employed, the same number of independent Fourier modes
are being utilized (splitting a survey box into $N$ slices decreases
the $k$-space density-of-states in each slice by a factor of $N$, but
this factor is recovered by averaging over the slices).  In order to
verify this point we re-analyzed a test case, dividing a broad
redshift range into several flat-sky slices such that the volume
contained within each slice was equal to the original curved-sky
volume (i.e.\ increasing the width of the flat-sky slice with
redshift).  We additionally varied the power spectum amplitude in each
slice in accordance with the cosmological growth factor.  We obtained
the final survey power spectrum by averaging the power spectra
measured for the individual slices with inverse-variance weighting.
The resulting power spectrum precision (recovered from the Monte Carlo
realizations as usual) agreed with our original single-slice analysis
to within 5 per cent.

\subsection{Angular selection functions}

In the present study we have not considered any effects due to complex
angular selection functions, assuming that our survey area is simply a
uniform box on the sky.  For a more complex survey geometry
(e.g.\ SDSS DR3), the observed galaxy power spectrum is smeared out
according to the Fourier transform of the observed sky area (and
correlations between adjacent power spectrum modes are induced).  This
convolution will smooth out features such as acoustic oscillations,
reducing their detectability, thus surveys with reasonably contiguous
geometries are required.  For example, the angular selection function
of the 2dFGRS (Percival et al.\ 2001) causes a smearing of the galaxy
power spectrum which is most significant on large scales (i.e.\ small
$k$) and almost a delta-function at $k \sim 0.1 \, h$ Mpc$^{-1}$
(Elgaroy, Gramman \& Lahav 2000).  Such a selection function would
therefore not seriously degrade measurement of the baryon
oscillations; detection of the turnover would be largely unaffected if
the width of the smearing was smaller than that of the turnover.

\subsection{Dust extinction}

Extinction by Galactic dust could affect the completeness of a survey
with a given magnitude limit, and also compromise the accuracy of the
photometric redshifts. This first problem has to be addressed for both
spectroscopic and imaging surveys alike, although the greatly
increased depth and areas covered by the imaging surveys imply that
this will be a larger problem relative to the small random errors.
This issue was discussed in detail by Efstathiou \& Moody (2001), and
was found to be unimportant for their results but would in general
affect the galaxy power spectrum on the largest scales.  For the
purposes of this paper we have assumed that these effects can be
overcome.  The second problem affects only the photometric redshift
surveys, and addressing this potential problem satisfactorily is
beyond the scope of this paper.  However, we note that in principle it
could be overcome if the photometric redshift algorithm could be
calibrated as a function of dust column density, for example using a
number of narrow spectroscopic surveys (`training sets') spanning a
range of Galactic extinction optical depths.

\section{Prospects for ongoing and future photometric redshift surveys}
\label{secrealsurv}

We now discuss our results in the context of existing or proposed
imaging surveys that may be utilized for photometric redshift studies.
Table \ref{tabfutsurv} provides a list of such surveys, covering both
optical and near infra-red wavebands.

\subsection{Photometric redshift estimation techniques}

A variety of techniques have been proposed in the literature for
derivation of photometric redshifts from multi-colour photometry.  The
photometric redshift performance depends on the method used, together
with a complex combination of the galaxy magnitude and redshift, the
filter set, the signal-to-noise ratio of the photometry, and the type
of galaxy spectrum (in general, redder objects yield more accurate
photometric redshifts).  A detailed simulation of this myriad of
factors is beyond the scope of this paper; however, some general
conclusions may be inferred using the photometric redshift accuracies
discussed in the literature.

The simplest photometric redshift techniques employ a limited set of
`template' spectra corresponding to local elliptical, spiral and
starburst galaxies (e.g.\ Hyper-Z; Bolzonella, Miralles \& Pello
2000).  These templates may be redshifted and fitted to observed
galaxy colours, deriving a likelihood distribution for the galaxy
redshift.  This approach has been successfully used in many cases
(e.g.\ the Hubble Deep Field North; Fernandez-Soto, Lanzetta \& Yahil
1999) and subject to various modifications such as the incorporation
of magnitude priors in a Bayesian framework (Benitez 2000) and the
iterative improvement of the initial templates (Budavari et
al.\ 2000).  Disadvantages of the method include the need for
spectro-photometric calibration over a wide wavelength range, the
difficulty of incorporating galaxy evolution with redshift, and the
failure of the template set to encompass all possible classes of
observed galaxy.

The availability of a {\it training set} -- spectroscopic redshifts
for a complete sub-sample of imaged galaxies -- is very helpful for
`tuning' the photometric-redshift technique.  In Table
\ref{tabfutcalib} we list a number of ongoing redshift surveys that
could potentially be used for this purpose.  In one possible
application of the training set, the galaxy redshift is expressed as a
polynomial in the colours, and the coefficients are fitted using the
training set (Connolly et al.\ 1995).  Further improvements are
possible if a different polynomial is adopted in each of a series of
cells in colour-space (Csabai et al.\ 2003).  Alternatively, an {\it
  artificial neural network} may be trained to deliver similar
information (Firth, Lahav \& Somerville 2003; Collister \& Lahav
2004).

\begin{table*}
\center
\caption{Existing and proposed photometric redshift imaging surveys;
we only list projects mapping $\approx 1000$ deg$^2$ or more.
For all future surveys the survey parameters are largely illustrative.}
\label{tabfutsurv}
\begin{tabular}{ccccc}
\hline
Survey & Waveband & Depth & Area (deg$^2$) & Start date \\
\hline
SDSS DR3 & ugriz & $r<20.8$ & 5282 & Released \\
UKIDSS & JHK & $K<18.5$ & 7500 & 2005 \\
KIDS & u'g'r'i'z' & $r'<24.2$ & 1700 & 2005 \\
VISTA Wide & YJHK & $K<19.5$, $Y<22.0$, $J<21.2$, $H<20.0$ & 3000 & 2006 \\
VISTA Atlas & JK & $K<18.2$, $J<20.2$ & 20000 & 2006 \\
Pan-STARRS & gRIZY & $R<27.2$ & 1200 & 2008 \\
CTIO DES & griz & $r<24.1$, $g<24.6$, $i<24.3$, $z<23.9$ & 5000 & 2009 \\
VISTA DarkCAM & u'g'r'i'z' & $r \la 25$ & $\sim$10000 & 2009 \\
LSST & rBgiz & $r<26.5$, $B<26.6$, $g<26.5$, $i<26$, $z<25$ & 15000 & $>$2012 \\
\hline
\end{tabular}
\end{table*}

\begin{table*}
\center
\caption{Existing and proposed spectroscopic redshift surveys.}
\label{tabfutcalib}
\begin{tabular}{cccccc}
\hline
Survey & Selection criteria & Area (deg$^2$) & No. galaxy redshifts & Status \\
\hline
CFRS & $I_{AB}<22.5$ & $\sim0.1$ & 591 & Released\\
CNOC-2 & $R<21.5$ & $1.5$ & 6200 & Released\\
COMBO17 (Photozs only) & $R<24$ & 0.25 & 10{,}000 & Released\\
SDSS DR3 & $r< 17.1$ & 4188 & 374767 & Released \\
VVDS CDFS (Le Fevre et al. 2004)
 & $I_{AB}< 24$ & 21x21.6 arcmin$^2$ & 1599 & Released \\
VVDS Deep & $I_{AB}< 24$ & 1.3 & 50,000 & Ongoing \\
VVDS Wide & $I_{AB}< 22.5$ & 16 & 100,000 &Ongoing \\
zCosmos & $I_{AB}<23$ & 2 & 90,000 & Ongoing\\
SDSS-2dF LRG
& $i<19.5$ plus colours for $0.4<z<0.8$& 300 & 10,000 & Ongoing\\
SDSS LRG & $i<19.2$ plus colours for $0.15<z<0.4$
& 5000 & 75000 & Ongoing \\
SDSS-II & $r<17.1$ & 10000 & $\sim 10^6$& Start 2005 \\
DEEP2 & $R_{AB}<24.1$ plus colours for $z>0.7$\footnote{Preselected
using BRI photometry} & 3.5 & 65,000 & Ongoing \\
KAOS & TBD & $\sim 1000$ & $\sim 10^6$ & Proposed\\
SKA & TBD & $\sim 30000$ & $\sim 10^9$ & Proposed \\
\hline
\end{tabular}
\end{table*}

\subsection{Existing surveys: SDSS}

The largest ongoing galaxy survey (Table \ref{tabfutsurv}) is the
SDSS.  Very recently, the {\it spectroscopic} component mapping
Luminous Red Galaxies has been utilized to obtain the first convincing
detection of the acoustic scale (Eisenstein et al.\ 2005).  The
inferred accuracy of the standard ruler in this study was $\approx
4\%$ and the rejection significance of $\Omega_{\rm b}/\Omega_{\rm m}
= 0$ was $3.4$-$\sigma$.  These results agree well with our own
simulation (see Section \ref{secpkerr}; we assume an expanded area of
$10{,}000$ deg$^2$) in which the average value of $P_{\rm rel} =
\exp{[-(\chi^2_{\rm no-wig} - \chi^2_{\rm wig-best})/2]}$ (Equation
\ref{eqprelwig}) over the Monte Carlo realizations is $0.035$
(i.e.\ $2.1$-$\sigma$ for the `model-independent' method -- although
$46\%$ of realizations perform better than 3-$\sigma$ -- with a full
$\Lambda$CDM fit expected to improve these figures as noted in Section
\ref{secbarfrac}).  Our simulated standard ruler precision is $2.5\%$
(the difference being explained by a scaling with survey area of
$\sqrt{A_\Omega}$).

Turning now to analyses of the SDSS {\it imaging} component, Csabai et
al.\ (2003) applied a variety of photometric redshift techniques to
the Early Data Release, determining an overall rms redshift scatter
$\delta z \approx 0.03$ for $r < 18$, rising to $\delta z \approx 0.1$
at $r \approx 21$, by which point systematic redshift discrepancies
due to large photometric errors have become important (i.e.\ the
effective magnitude limit for reliable application of photometric
redshifts may be $r_{\rm lim} \approx 20$). The VVDS survey databases
would constitute a suitable spectroscopic calibration set (see Table
\ref{tabfutcalib}) for photometric redshift techniques that require
it.  As discussed in Section \ref{secphotozerr}, the inevitable
presence of a small fraction of interlopers with significantly larger
redshift errors does not compromise the scientific results.

Combining these results with our Figures \ref{figwigconfphoto} and
\ref{figturnconfphoto} we conclude that a photometric redshift
analysis of the entire SDSS imaging database may not succeed in
detecting features in the galaxy power spectrum in the
model-independent manner discussed in Sections \ref{secwig} and
\ref{secturn}. For example even if $\sigma_0 = 0.03$ and $r < 20.5$
were possible then Figures \ref{figwigconfphoto} and
\ref{figturnconfphoto} would imply a wiggle `detection' with 90 per
cent confidence and a turnover `detection' with 80 per cent
confidence.  However the prospects for using the $\Lambda$CDM fit in
Figure~\ref{figconf_obom} are better, implying a detection of
$\Omega_{\rm b}/\Omega_{\rm m} \ne 0$ with a significance of
8-$\sigma$ (for fixed values of $h$ and $n_{\rm s}$).

However, it is clear that certain sub-classes of galaxy perform
significantly better regarding photometric redshifts: SDSS red
galaxies yield an accuracy twice that of blue galaxies (Csabai et
al.\ 2003), and Luminous Red Galaxies (LRGs) permit a rms redshift
precision of $\sigma_0 = 0.02$ for $z < 0.55$ (Padmanabhan et
al.\ 2005).  In addition, these LRGs inhabit massive dark matter
haloes and are consequently biased tracers of the large-scale
structure (we assume a linear bias factor $b = 2$).  The resulting
amplification of the clustering strength implies that a lower space
density is required to yield a given power spectrum accuracy, although
in some models of galaxy clustering the amplitude of acoustic
oscillations is diluted for the most massive galaxies (e.g.\ Peacock
\& Smith 2000).

The optimal imaging approach is therefore to analyze an LRG {\it
  photometric redshift catalogue} (a suitable spectroscopic training
set is the SDSS-2dF LRG redshift survey, see Padmanabhan et
al.\ 2005).  We simulated such a catalogue using a redshift interval
$0.2 < z < 0.7$, assuming a Gaussian redshift distribution peaking at
$z = 0.45$ of standard deviation $z = 0.1$. We supposed LRGs could be
selected from the photometric data with a surface density $\Sigma_0 =
100$ deg$^{-2}$ and photometric redshifts measured with an accuracy
$\sigma_0 = 0.02$ (i.e.\ $\delta z = \sigma_0(1+z_{\rm eff}) \approx
0.03$; see Padmanabhan et al.\ 2005).  In addition, LRGs are assigned
a linear bias factor $b = 2$. Some of these assumptions may be
optimistic, in particular Padmanabhan et al.\ note that the
photometric redshift accuracy degrades for $z > 0.55$.  We quantified
the confidence of detection of acoustic oscillations for a $10{,}000$
deg$^2$ survey.  The average value of $P_{\rm rel}$ is $0.056$,
corresponding to a detection slightly less confident than but
comparable to the LRG spectroscopic database, with $28\%$ of
realizations possessed a `model-independent' confidence exceeding
3-$\sigma$.  The simulated accuracy of the standard ruler measurement
is $3.8\%$.  For the turnover analysis, our simulated detection
confidence is low: we derive an average value of $P_{\rm rel} = 0.33$,
with an accuracy for the turnover scale of $27\%$.  Again, individual
realizations may perform significantly better.

We conclude that the SDSS imaging dataset has the potential to yield a
marginal model-independent detection of acoustic oscillations using a
sub-sample of red galaxies.  However, we caution that the
currently-available third SDSS data release (DR3), which covers a sky
area $\approx 5000$ deg$^2$, possesses a complicated angular window
function which renders this experiment more difficult: as discussed in
Section \ref{secapprox}, the observed power spectrum is a convolution
of the underlying power spectrum and the survey window function; if
this latter possesses a substantial width in Fourier space then the
oscillatory signal will be smoothed and consequently harder to detect.

\subsection{Future surveys}

We now turn our attention to future surveys.  We first note that the
availability of {\it near infra-red imaging} to appropriate depths is
extremely valuable for galaxies with redshifts $z > 0.4$.  In this
range, Bolzonella, Miralles \& Pello (2000) derived a factor of 2
improvement in redshift accuracy when $JHK$ photometry was added to
the standard optical wavebands.  The combination of the future UKIDSS
infra-red data (see Table \ref{tabfutsurv}) with the ongoing SDSS
optical survey will therefore be very powerful, and should permit
detection of the acoustic oscillations using the complete galaxy
population rather than special sub-classes. For example, if the
photometric error were halved from $\sigma_0 = 0.03$ to $\sigma_0 =
0.015$ (assuming a magnitude threshold $r = 20.5$) then the
model-independent wiggle detection limits in Figure
\ref{figwigconfphoto} show a detection at 99 per cent confidence (as
opposed to just over 90 per cent without the infra-red data).

It is clear from Figure \ref{figwigconfphoto} that in order to deliver
a {\it high-significance} measurement of the acoustic peaks using
photometric redshifts (together with a significant detection of the
power spectrum turnover), a much deeper optical database is required
($r \sim 24$) over an area of several thousand deg$^2$.  Such a
catalogue may first be provided by the CTIO Dark Energy Survey (DES;
see Table \ref{tabfutsurv}).  In this case, a photometric-redshift
precision $\sigma_0 = 0.05$ suffices for delineation of the acoustic
oscillations and turnover, a redshift accuracy which has been achieved
in existing analyses of the Hubble Deep Field (e.g.\ Fernandez-Soto et
al.\ 1999). Furthermore, judged solely by the accuracy with which the
power spectrum features can be mapped out, this photometric-redshift
approach will be competitive when compared to the spectroscopic
redshift surveys which will be contemporary to the DES (e.g.\ the KAOS
proposal) -- although we note that in terms of measuring the dark
energy parameters, a spectroscopic survey yields critical additional
information in the radial direction (namely, the Hubble constant at
high redshift) which is forfeited by the photometric-redshift
approach.

The `ultimate' photometric-redshift survey would cover an area
$A_\Omega$ approximating the whole sky to a magnitude depth $r \sim
26$; surveys with the LSST will approach these specifications (see
Table \ref{tabfutsurv}).  If we assume that the redshift accuracy
cannot exceed the limit $\sigma_0 = 0.01$, in this best case a
competing spectroscopic survey would need to cover $A_\Omega/4 \sim
10000$ deg$^2$ to produce comparable power spectrum constraints
(i.e.\ trace the same number of Fourier structure modes).  In optical
wavebands, spectrographs with fields-of-view greatly exceeding $1$
deg$^2$ are prohibitively expensive, but we note that redshift surveys
for neutral hydrogen using a next-generation radio telescope such as
the Square Kilometre Array would become competitive here (Blake et
al.\ 2004), provided that such a telescope was designed with an
instantaneous field-of-view of order $100$ deg$^2$ at frequency $1.4$
GHz.

We note that ultimately the matter power spectrum derived from
gravitational lensing will circumvent the assumptions about bias
necessary for the interpretation of galaxy surveys.  Of course,
lensing investigations can be performed with the identical imaging
datasets discussed here, and indeed the application of photometric
redshifts will form an important part of that interpretation.
However, due to the unknown intrinsic shapes of galaxies, many
galaxies have to be averaged to obtain a cosmic shear
signal. Therefore we will always be able to obtain constraints from
the galaxy power spectrum that are vastly tighter than those inferred
from the lensing power spectrum, albeit less reliable owing to the
caveats regarding bias.  Thus galaxy surveys are the best place to
look for the first glimpses of any exciting new physics revealed by
studies of large-scale structure.

\section{Conclusions}

We have used Monte Carlo techniques to estimate the measurement
precision of the galaxy power spectrum achievable using photometric
redshift imaging surveys with a variety of magnitude depths and
photometric redshift accuracies.  We have focussed on two main areas:
\\
$\bullet$ The `model-independent' measurement of specific features in
the galaxy power spectrum: the acoustic oscillations and the turnover.
In particular, we have carefully quantified the statistical confidence
with which these features may be detected, together with the accuracy
with which the acoustic and turnover scales can be inferred.
\\
$\bullet$ The assumption of the $\Lambda$CDM paradigm to place tighter
constraints on the baryon fraction $\Omega_{\rm b}/\Omega_{\rm m}$
from galaxy surveys alone, and an evaluation of the additional
constraints on the running of the spectral index of the primordial
power spectrum as increasingly powerful galaxy surveys are combined
with WMAP.

We summarize our general findings as follows:
\\
$\bullet$ On linear-regime scales, a large photometric redshift survey
can provide competitive power spectrum measurements when compared to
contemporaneous spectroscopic surveys.  For example, given a magnitude
threshold $r_{\rm lim} = 23$, a $10{,}000$ deg$^2$ photometric
redshift survey with error parameter $\sigma_0 = 0.03$ results in a
confidence of detecting the acoustic oscillations that is comparable
to a spectroscopic survey covering $1000$ deg$^2$.  (Albeit with the
loss of radial information that is helpful for constraining dark
energy models).
\\
$\bullet$ We compare various different definitions of `detection of
acoustic oscillations', the most optimistic of which is a $\Lambda$CDM
fit to the data (i.e.\ derivation of a baryon fraction significantly
exceeding zero) which (for fixed values of $h$ and $n_{\rm s}$) yields
detections with statistical significance approximately a factor of
four $\sigma$'s greater than our default `model-independent' fit
using no shape information.
\\
$\bullet$ Concerning the power spectrum turnover, the relevant scales
are sufficiently large that photometric redshifts with precision
better than $\sigma_0 = 0.04$ retain information equivalent to
spectroscopic surveys.  An imaging depth $r \sim 24$ is required to
deliver a 3-$\sigma$ detection confidence.
\\
$\bullet$ Our results are robust against more complex photometric
redshift error distributions such as double Gaussian models, if our
parameter $\sigma_0$ is interpreted as the overall rms error.

In terms of realistic ongoing and future surveys:
\\
$\bullet$ The SDSS has already yielded a convincing detection of
acoustic features in the spectroscopic survey of Luminous Red
Galaxies.  A marginal `model-independent' detection may also follow
from this sample.
\\
$\bullet$ Analysis of an LRG photometric redshift database selected
directly from the imaging data may yield a measurement of comparable
precision.
\\
$\bullet$ Assuming a $\Lambda$CDM model and calculating the number of
standard deviations by which $\Omega_{\rm b}/\Omega_{\rm m}$
(marginalized over the shape parameter $\Omega_{\rm m} h$) exceeds
zero shows that, for our simulated full SDSS spectroscopic survey the
baryons would be detected at $3.5$-$\sigma$, whereas for a photometric
redshift survey to magnitude depth $r < 21$ with $\delta z = 0.03
(1+z)$ the detection level is 9-$\sigma$.  We note that these analyses
assume a conservative specification of the maximum wavenumber $k_{\rm
  max}$ for which a linear power spectrum is fitted.
\\
$\bullet$ A high-significance analysis of the acoustic oscillations
from photometric redshifts, and a model-independent detection of the
power spectrum `turnover', requires a significantly deeper optical
database ($r \sim 24$) over an area of several thousand deg$^2$.

\section*{Acknowledgments}

CB acknowledges Karl Glazebrook for an invaluable collaboration
developing simulations of experiments measuring the wiggles, and for
useful comments on a draft of this paper.  CB is also grateful to the
Australian Research Council for financial support, and acknowledges
current funding from the Izaak Walton Killam Memorial Fund for
Advanced Studies and the Canadian Institute for Theoretical
Astrophysics.  We thank Adrian Collister, George Efstathiou, Ofer
Lahav and David Woods for helpful comments.  SLB thanks the Royal
Society for support in the form of a University Research
Fellowship. SLB thanks Antony Lewis for various improvements to
CosmoMC. CosmoMC calculations were performed on the UK National
Cosmology Supercomputer Center funded by PPARC, HEFCE and Silicon
Graphics / Cray Research.  We thank Andy Taylor and Konrad Kuijken for
advising on the VISTA DarkCAM and KIDS planned survey parameters,
respectively.  We thank Berkeley for hospitality while this work was
being finished.

\end{document}